\documentclass[%
 reprint,
superscriptaddress,
nofootinbib,
nobibnotes,
 amsmath,amssymb,
 aps,
]{revtex4-2}

\usepackage{graphicx}
\usepackage{dcolumn}
\usepackage{bm}

\usepackage[caption=false]{subfig}
\usepackage{amssymb}
\usepackage{amsmath}
\usepackage{commath}
\usepackage{graphicx,bm}
\usepackage{verbatim}
\usepackage{xcolor}
\usepackage{xr-hyper}
\usepackage{url}
\usepackage{xurl}
\usepackage{hyperref}

\usepackage{soul}

\graphicspath{{Figures/}} 

\begin{document}

\preprint{APS/123-QED}

\title{The different structure of economic ecosystems at the scales of companies and countries}

\author{Dario Laudati}
 \affiliation{Dipartimento di Fisica, Sapienza Universit\`a di Roma, 00185, Rome, Italy}
 \author{Manuel S. Mariani}
 \affiliation{Institute of Fundamental and Frontier Sciences, University of Electronic Science and Technology of China, Chengdu 610054, PR China}
\affiliation{URPP Social Networks, Universität Zürich, 8050 Zürich, Switzerland}
\author{Luciano Pietronero}%
\affiliation{%
 Centro Ricerche Enrico Fermi, 00184, Rome, Italy
}%
\author{Andrea Zaccaria}%
 \email{andrea.zaccaria@cnr.it}
\affiliation{%
 Istituto dei Sistemi Complessi, UOS Sapienza, CNR, 00185, Rome, Italy
}%

\date{\today}



\begin{abstract}
A key element to understand complex systems is the relationship between the spatial scale of investigation and the structure of the interrelation among its elements. When it comes to economic systems, it is now well-known that the country-product bipartite network exhibits a nested structure, which is the foundation of different algorithms that have been used to scientifically investigate countries' development and forecast national economic growth. Changing the subject from countries to companies, a significantly different scenario emerges. Through the analysis of a unique dataset of Italian firms’ exports and a worldwide dataset comprising countries' exports, here we find that, while a globally nested structure is observed at the country level, a local, in-block nested structure emerges at the level of firms. Remarkably, this in-block nestedness is statistically significant with respect to suitable null models and the algorithmic partitions of products into blocks have a high correspondence with exogenous product classifications. These findings lay a solid foundation for developing a scientific approach based on the physics of complex systems to the analysis of companies, which has been lacking until now.
\end{abstract}

\maketitle

Understanding the structure of interactions in a complex system is a fundamental issue \cite{anderson1972more, pietronero2008complexity}, since the structure affects the system's function~\cite{lynn2019physics,morone2019symmetry} and its resilience against diverse perturbations~\cite{rohr2014structural,dominguez2015ranking,morone2019k,arese2020diversity}. Yet, interactions can be bounded by different kinds of constraints~\cite{lewinsohn2006structure}. When this is the case, understanding the structure and dynamics of interactions requires to identify clear boundaries that separate an ecosystem from its surroundings.
While this idea and the resulting methods~\cite{lewinsohn2006structure,kojaku2017finding,sole2018revealing,mariani2019nestedness} have found promising initial applications in ecological~\cite{flores2013multi,lampo2021hybrid} and social networks~\cite{palazzi2019online,palazzi2021ecological}, they have not yet been applied to economic systems where actors produce and export products. As for these systems, most studies assume that the ecosystem where a country operates is the entire world~\cite{serrano2003topology,garlaschelli2005structure,saracco2015randomizing}: in principle, each country competes with all the others, and all products are considered.
To uncover the complexity of countries' export structure, the world trade web is often represented as a bipartite network where countries and products constitute the nodes of the two layers \cite{hidalgo2009building,tacchella2012new}. At this global scale, a peculiar property emerges: nestedness \cite{mariani2019nestedness}. Well-known in ecology, in this context nestedness means that developed countries are highly diversified and produce all kinds of products, while poor countries only produce few ubiquitous products.
This empirical observation led to the development of Economic Complexity, an interdisciplinary approach which applies methods from statistical physics and network science to uncover the determinants of country development \cite{hidalgo2009building,tacchella2012new}. 
Notably, a predictive approach based on nestedness \cite{tacchella2012new} is able to forecast GDP growth with a significant improvement over the IMF projections \cite{cristelli2017predictability, tacchella2018dynamical}.

Despite these remarkable achievements, a fundamental question remains still open: given that the export of countries is nothing more than the result of the production of individual companies at national level, can the Economic Complexity approach be extended to the scale of companies?

Answering this question has been hindered by two main factors. First, data scarcity: export data at the company level is extremely sensitive in terms of privacy policy and is much less homogeneous with respect to the harmonized data about the international trade. Second, and more importantly, the networks of countries and companies may have different structures and, regarding firms, the correct ecosystem to consider is still unknown.

Thanks to our collaboration with the Italian National Institute of Statistics (IT ISTAT), we could overcome these limitations and access a unique dataset of Italian firms' export records. The products are coded in the same way as previously-analysed datasets of countries' exports, which enables a direct comparison of the structures of the country-product and company-product ecosystems.

Building on this dataset, we apply algorithms to statistically validate the presence of modularity, nestedness, and in-block nestedness~\cite{sole2018revealing,palazzi2021ecological} to both the country-product and the company-product networks. We find that the same level of nestedness which is present at the country scale is absent when one looks at a national economy of companies as a whole, but re-emerges at the local level, once that the modular structure of the company-product network is considered. As a result of these structural differences, the ranking algorithms developed to evaluate countries' competitiveness do not work properly when applied to the network of companies as a whole. At the same time, the detected in-block nestedness of the company-product network opens up the possibility to apply the Economic Complexity framework also at the company level, provided that the proper locally-nested ecosystems are considered: the company, its competitors, and the products they compete on.

\section*{\label{sec:Results}Results}


\subsection*{Same ranking algorithm, different conclusions}


Economic Complexity algorithms were originally designed to evaluate the competitiveness of countries and the complexity of products from the structure of the country-product network~\cite{hidalgo2009building,tacchella2012new}.
We begin by showing that state-of-the-art Economic Complexity algorithms are inadequate to capture the complexity of products and the competitiveness of firms from the structure of the company-product network \cite{bruno2018colombian}. 
To demonstrate this point, we apply the Fitness-Complexity algorithm \cite{tacchella2012new,pugliese2016convergence} (see Methods for the mathematical formulation) to both the country-product and the company-product networks. In this way, we obtain two different evaluations of the same quantity: the Complexity of products. To assess whether the obtained Complexity scores are good proxies for the economic value of a product, we compare the Complexity rankings with those obtained according to the $\log$PRODY index (see Methods), an external monetary metric that measures the sophistication of products from the GDP of the exporting countries \cite{hausmann2007you,angelini2017complex}. We expect that a reasonable measure of Complexity should exhibit a good correlation with $\log$PRODY.


\begin{figure*}[t]
  \includegraphics[width=.49\linewidth]{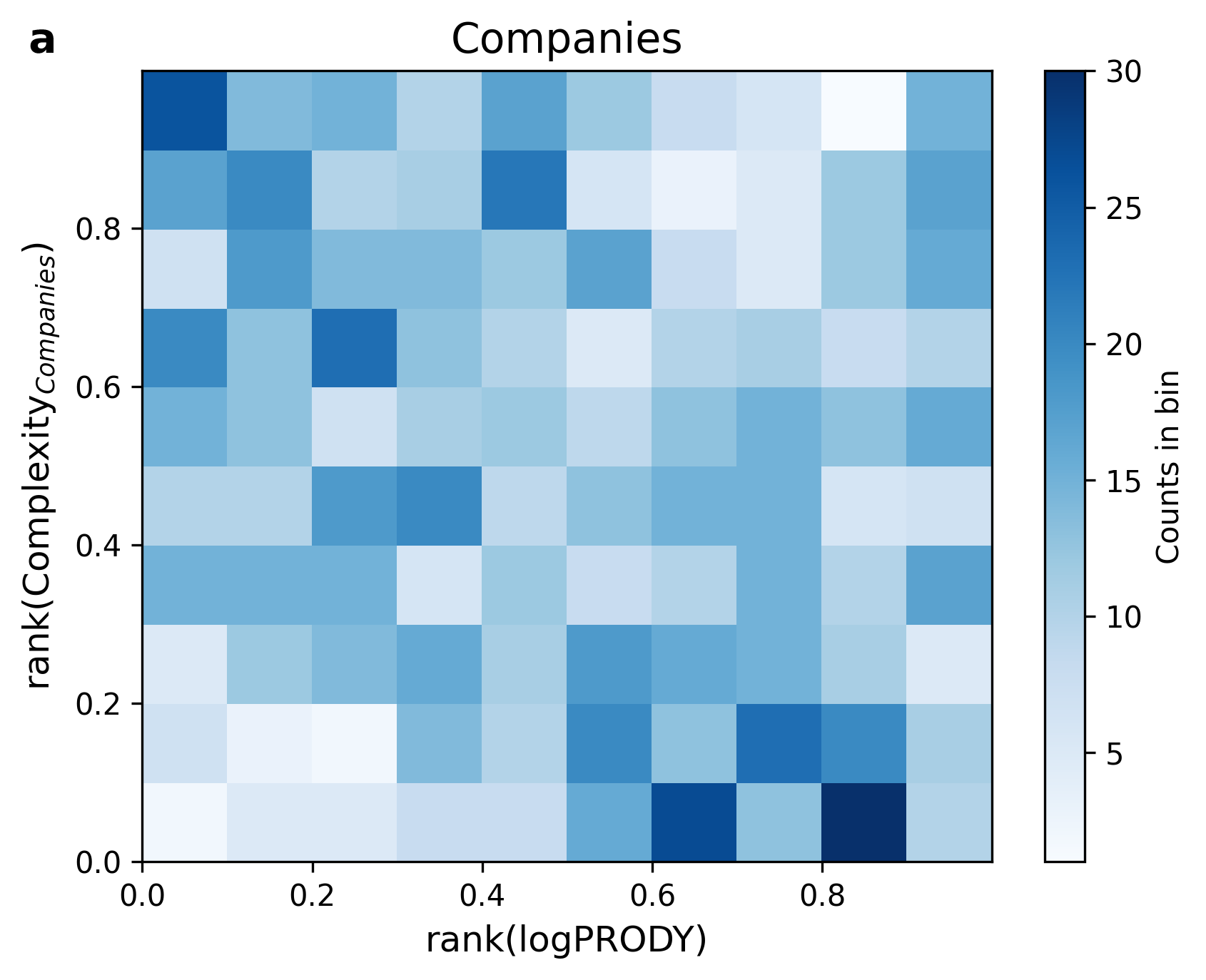}%
  \includegraphics[width=.49\linewidth]{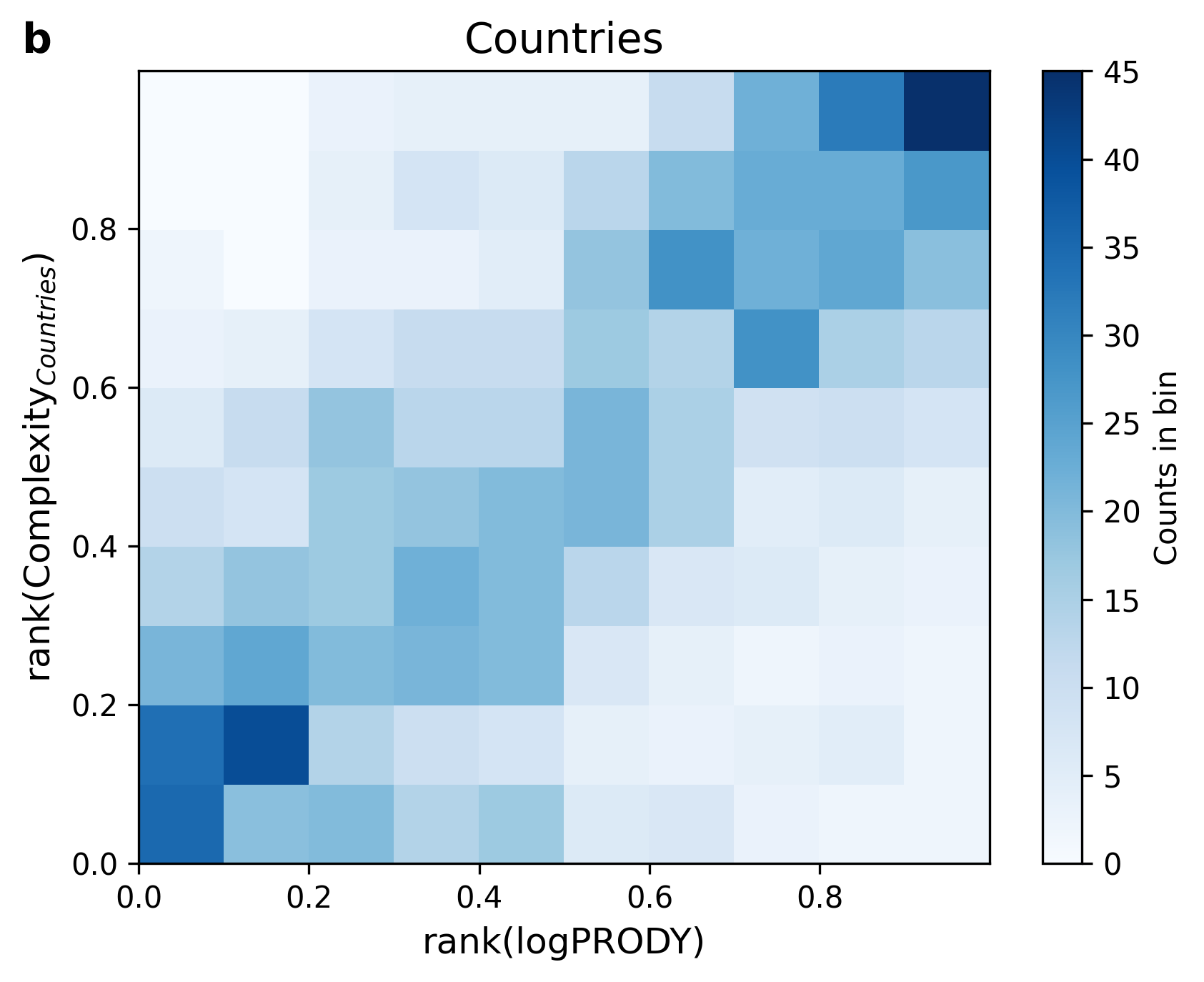}%
\caption{\textbf{Same ranking algorithm, different product Complexity.} The 2d histograms compare the product rankings obtained with the Fitness-Complexity algorithm at the company (left) and the country (right) level with those computed using the logPRODY index, an economic-based measure of sophistication. Points are grouped in bins of size 0.1, with rankings being normalised between 0 and 1. In the case of similar classifications, an accumulation of points around the secondary diagonal should be observed. We find that the Complexity of products is correlated with the logPRODY index when it is extracted by applying the Fitness-Complexity algorithm to the country-product network (b), but not when extracted from the company-product network (a). A possible explanation lies in the different structures of the two networks.}
\label{3}
\end{figure*}

A good agreement between Complexity and  $\log$PRODY is only observed when the Complexity score is obtained by applying the Fitness-Complexity algorithm to the country-product network (Spearman’s correlation coefficient $\rho=0.642$), but not when the same algorithm is applied to the company-product network ($\rho=-0.224$, see Fig.~\ref{3}).
A similar conclusion is reached by comparing the countries' and companies' (extensive) Fitness scores with their export volumes, which can be interpreted as proxies for their competitiveness. We only observe a high correlation for countries ($\rho=0.887$), but not for companies ($\rho=0.378$) - see Supplementary Sect. II and \cite{bruno2018colombian} for a comparison with the degree. These results indicate that when applied to the company-product network, the Fitness-Complexity algorithm does not accurately estimate the economic value of products and the competitiveness of the exporters. It comes therefore natural to wonder why the Fitness-Complexity algorithm fails in the company-product network.



A possible answer lies in the different structures of the company-product and country-product networks. The Fitness-Complexity algorithm builds on the premise that competitive countries tend to diversify their export baskets as much as possible, given their available capabilities~\cite{hausmann2011network,cristelli2013measuring,saracco2015innovation}.
This hypothesis is motivated by the globally nested structure of the country-product network~\cite{tacchella2012new,mariani2019nestedness}: the most diversified countries export all kinds of products, whereas the products exported by more specialised countries are typically exported by the diversified ones as well. Drawing a parallel with ecology, a more diversified export basket might increase the robustness of a country's economy with respect to adverse external events~\cite{dominguez2015ranking,mariani2015measuring}.
A similar argument might, in principle, apply to firms as well, and prior works have associated the diversification of a firm's activities with its economic performance~\cite{miller2004firms, jose1986contributions, michel1984does}. Yet, other studies emphasised the importance for a firm to diversify within its core set of capabilities, avoiding unrelated activities~\cite{palepu1985diversification,pugliese2019coherent, christensen1981corporate, valvano2003diversification, bruno2018colombian,kim2021technological}. These works suggest that a globally nested structure might not be found, and that the company-product network might be instead partitioned into specialised blocks, which would disagree with the basic premise of the Fitness-Complexity algorithm.

\subsection*{The different role of modularity}


The ultimate test of these conjectures lies in the empirical data. To identify potential differences in the structure of the two economic systems, we apply a modularity maximisation algorithm (BRIM, see Methods for more details) to both.
By only looking at the modularity scores of the two networks, one may naively conclude that both exhibit a pronounced modular structure ($Q=0.218, p<0.01$ for the country-product network, $Q=0.512, p<0.01$ for the company-product network, where the $p$-values have been obtained with the BiCM null model, see Methods). 

Yet there are two substantial differences between the two modular patterns.
First, the detected partitions are much noisier in the country-product network than in the company-product network (see Supplementary Fig. S2 for a visual comparison).
More specifically, the blocks in the country-product network contain only 50\% of the links, whereas in the company-product network they contain more than 70\% of the links.

Second, the interpretation of the detected blocks is radically different in the two systems.
To interpret the detected blocks, we investigate their sector composition. To this end, we compare the modularity-detected partition of products with the ones corresponding to the 21 sections of the official export classification, that is the Harmonized System (HS), 1992 edition (see Supplementary Sect. V for a detailed description of this classification). The basic idea is that, since the HS sections represent homogeneous categories of products, then coherent (specialised) partitions should show a substantial degree of relatedness, that is, similar products should co-occur in the same blocks. Conversely, heterogeneous (diversified) partitions should show a low degree of relatedness. To ensure the robustness of our conclusions, we perform the comparison between HS sections and the partitions extracted by several different community detection algorithms (BRIM and BRIM$^2$ \cite{platig2016bipartite}, BiLouvain \cite{blondel2008fast}, and IBN \cite{sole2018revealing} - see Methods for more details): robust results should not depend on the particular algorithm employed, as long as it provides a reasonable partition. The similarity between the partitions is measured using the Adjusted Mutual Information \cite{vinh2010information} (AMI, see Methods for the mathematical definition). We emphasise that this analysis does not aim to evaluate the detected partitions~\cite{peel2017ground}, but only to provide a robust interpretation of the detected modules.

\begin{figure*}[t]
\centerline{\includegraphics[width=.49\linewidth]{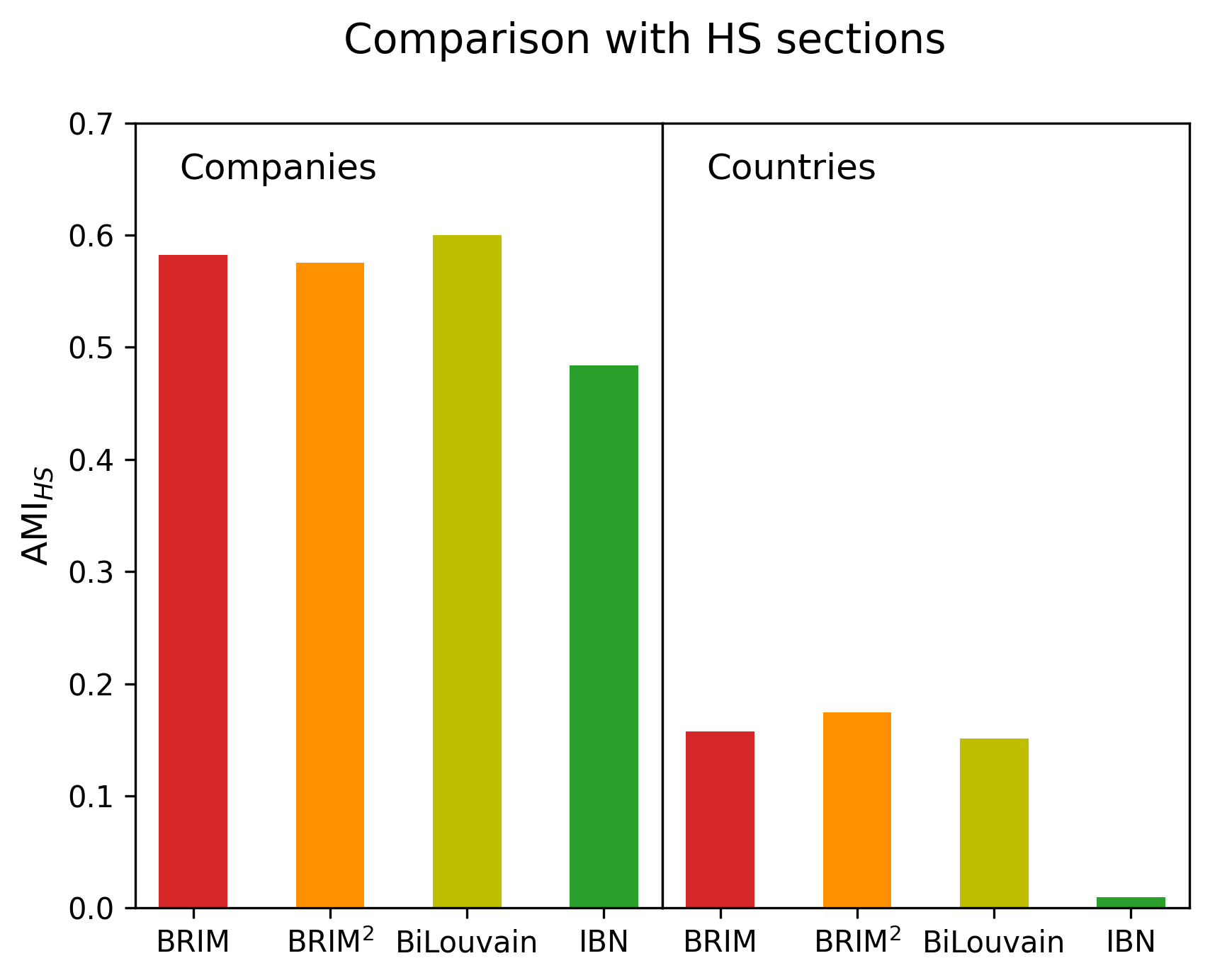}}
\caption{\textbf{A comparison between the detected product partitions and the HS categorisation} for the countries' and firms' ecosystems. Similarity between classifications is measured through the Adjusted Mutual Information (AMI), which is based on the idea that if two partitions are similar, one needs very little information to infer one partition given the other (see Methods for more details). In the case of companies, the division of products closely resembles the homogeneous classification provided by the HS System (high AMI), while the same does not hold true for countries, where the identified blocks are characterised by a pronounced heterogeneity (low AMI). 
}
\label{4}
\end{figure*}

We find that the AMI is significantly larger in the firm-product than in the country-product network. 
For example, by using the BRIM algorithm, the AMI is 270\% larger in the company-product network than in the country-product network. Qualitatively similar results hold for other algorithms (see Fig.~\ref{4}).
This set of results indicate that companies are genuinely specialised entities that mostly focus on homogeneous groups of products. By contrast, countries do not specialise in confined groups of similar products, as proven by the high heterogeneity of the detected blocks: developed countries diversify their production~\cite{tacchella2012new}. It can be shown that the significant degree of modularity observed in the country-product network can be explained by the countries' diversification patterns (see Supplementary Sect. III B for a detailed discussion).

\begin{figure*}[t]
\includegraphics[width=.49\linewidth]{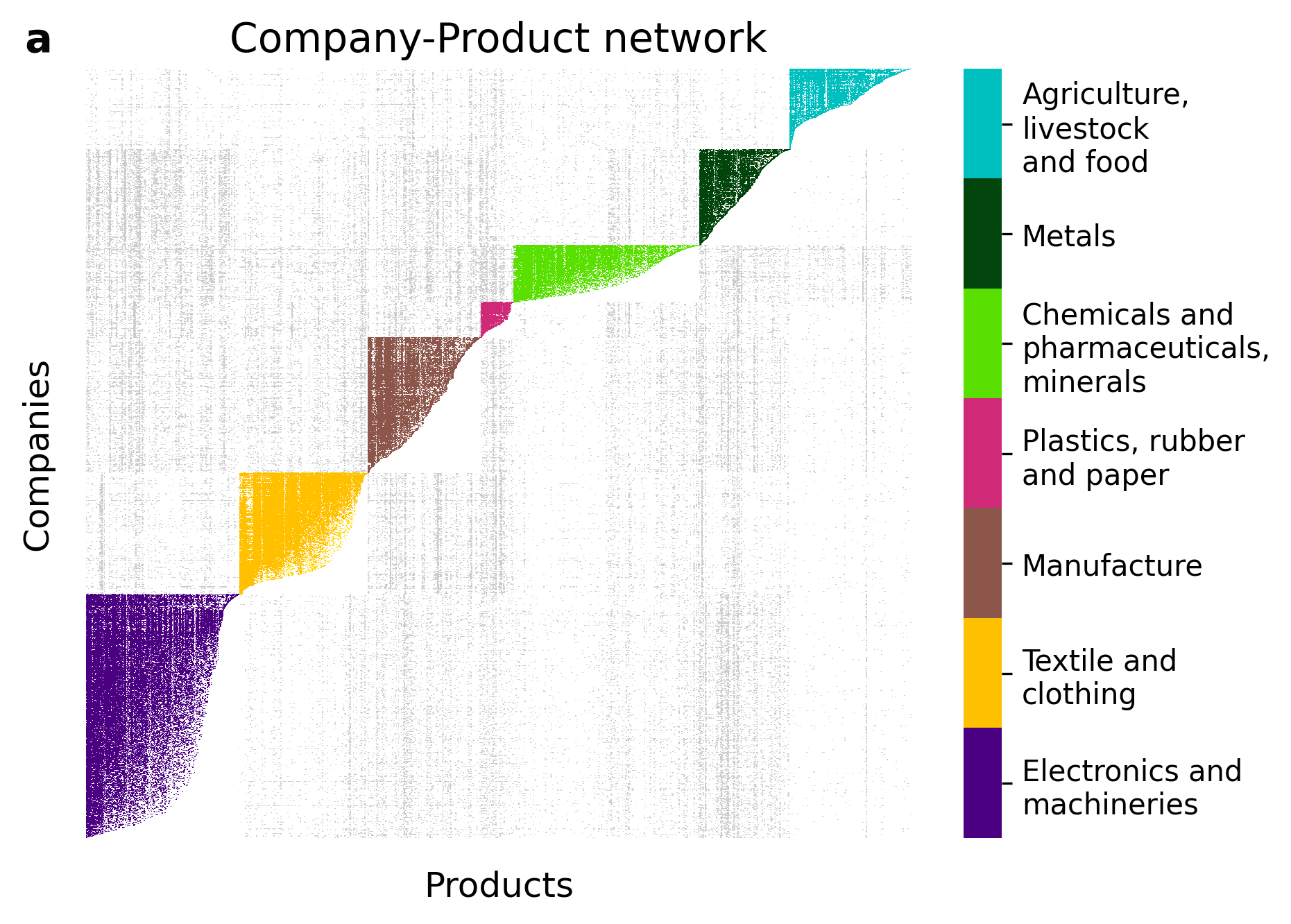}%
  \includegraphics[width=.49\linewidth]{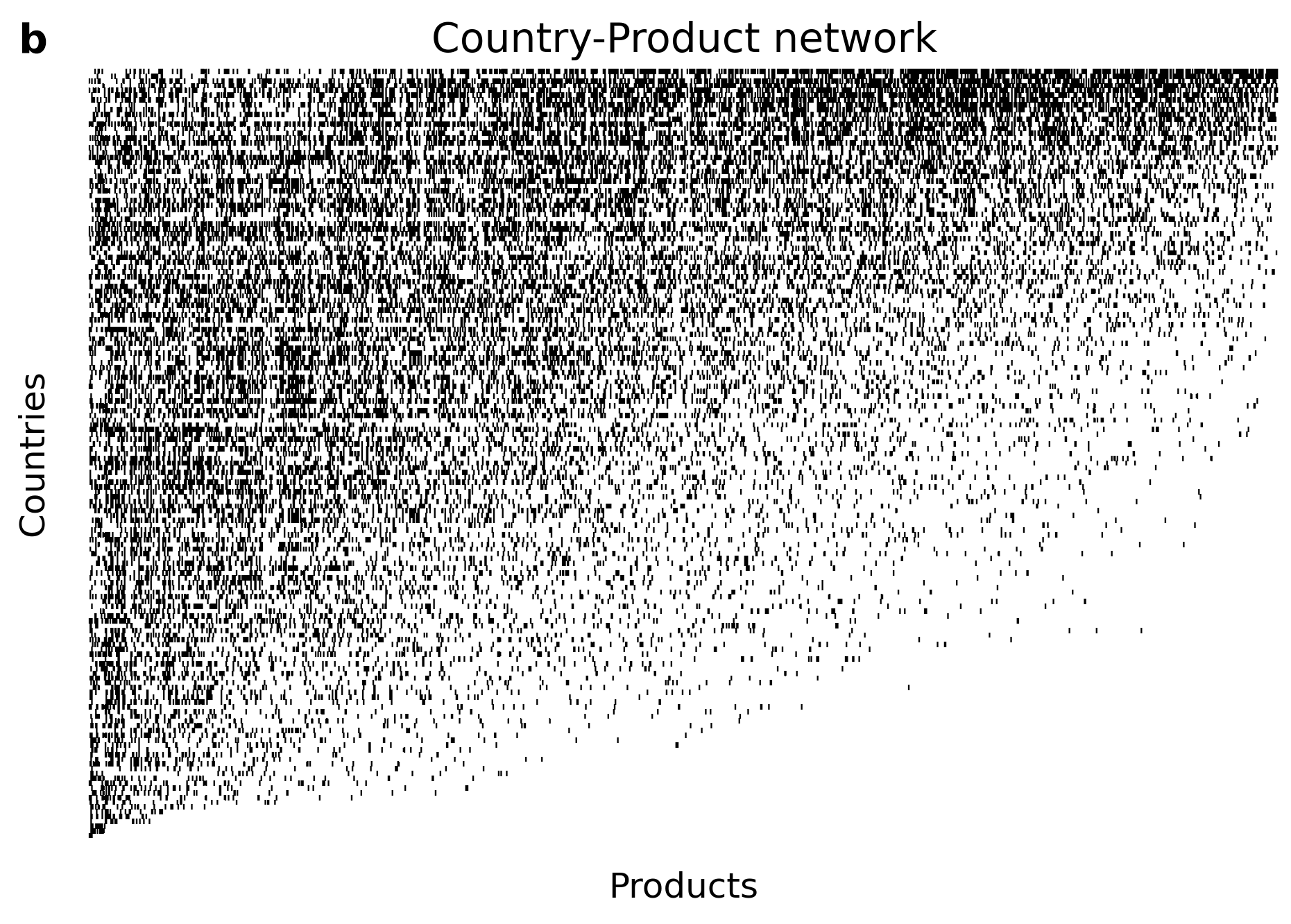}%
\caption{\textbf{The different structure of the bipartite export network at company and country level.} While the country-product network exhibits a globally-nested structure, the firm-product network can be partitioned into blocks that exhibit an internal nested structure. (a) Firm-product export network. Within each module detected by the BRIM modularity maximisation algorithm (coloured blocks), rows and columns have been sorted according to the Fitness-Complexity algorithm. The colours of each module reflect the economic sector represented by the majority of products in the module. (b) Country-product export network, where rows and columns have been sorted according to the Fitness-Complexity algorithm.\\
While for countries the proper ecosystem is the whole world, in the case of companies local ecosystems in line with the intuitive sectoral divisions emerge.
}
\label{1}
\end{figure*}

These results lead to the investigation of the internal structure of the detected company-product blocks. In particular, the interesting point to evaluate is whether there is a resemblance between the structure of these blocks and the global structure that characterises the country-product network. Such evidence would support the idea that the detected blocks act as boundaries that constrain the companies' ability to diversify. To this end, we apply the Fitness-Complexity algorithm to the BRIM blocks, and we use the rankings to order the company-product matrices. The result is depicted in Fig. \ref{1}a. Besides a good agreement with the industrial sectors, the blocks identified in the company-product network display another very interesting feature: they exhibit an internally nested structure. This property will be deeply investigated in the next section.

In the light of these results, an initial characterisation for the two economic ecosystems can be outlined. For countries, nestedness is the dominant property, whereas modularity (although statistically significant) emerges only as a second-order feature, essentially determined by the countries' diversification. For companies, modularity is the dominant property, whereas nestedness is relegated within blocks, as a local property.
We can then argue that the reason why the Fitness-Complexity algorithm, if applied globally to the company-product network, misestimates the sophistication of products and the competitiveness of companies is that it neglects the block structure of the network.

\subsection*{Local nestedness in the firms' ecosystem}

A full validation of the previous characterisation of the two systems requires the deployment of methods that can disentangle the role of nestedness and modularity~\cite{sole2018revealing}.
Specifically, to prove the claim that nestedness is a local (global) property in the firms' (countries') ecosystems, we implement a recent method to rigorously determine whether a network can be partitioned into blocks with an internal nested structure~\cite{sole2018revealing}. This method relies on a quality function -- referred to as \textit{in-block nestedness}, $\mathcal{I}$ (see Methods for the mathematical definition) -- and requires to optimise the in-block nestedness function and to compare its optimal value, $\mathcal{I}^{*}$, against the value of the same function for a single-block partition, which we refer to as $\mathcal{N}$ (see Methods). Large values of the ratio $\mathcal{I}^{*}/\mathcal{N}$ indicate that nestedness is a local property, while networks where nestedness is a global property exhibit $\mathcal{I}^{*}\simeq\mathcal{N}$~\cite{sole2018revealing,palazzi2021ecological}.

\begin{figure*}[t]
  \includegraphics[width=.49\linewidth]{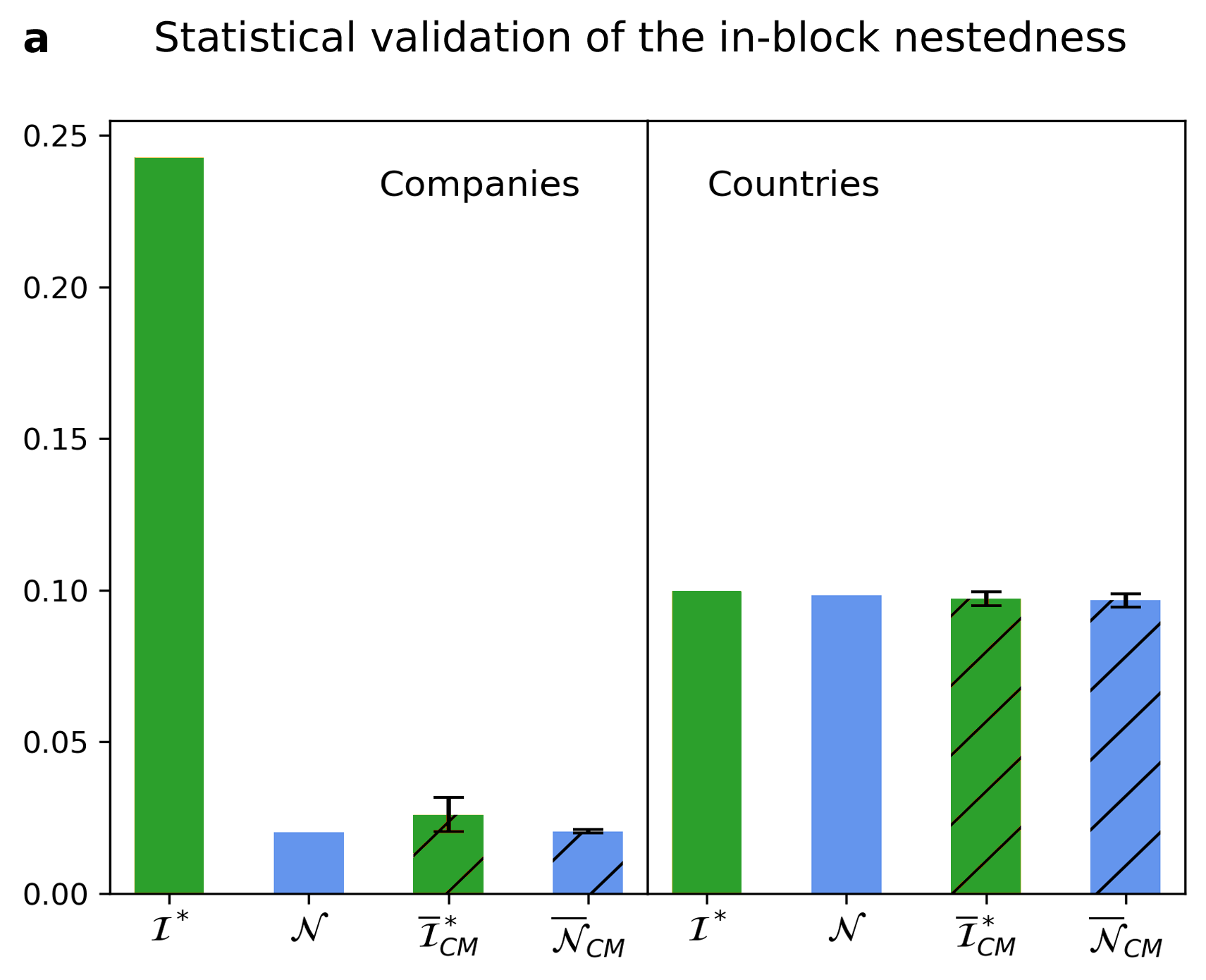}%
  \includegraphics[width=.49\linewidth]{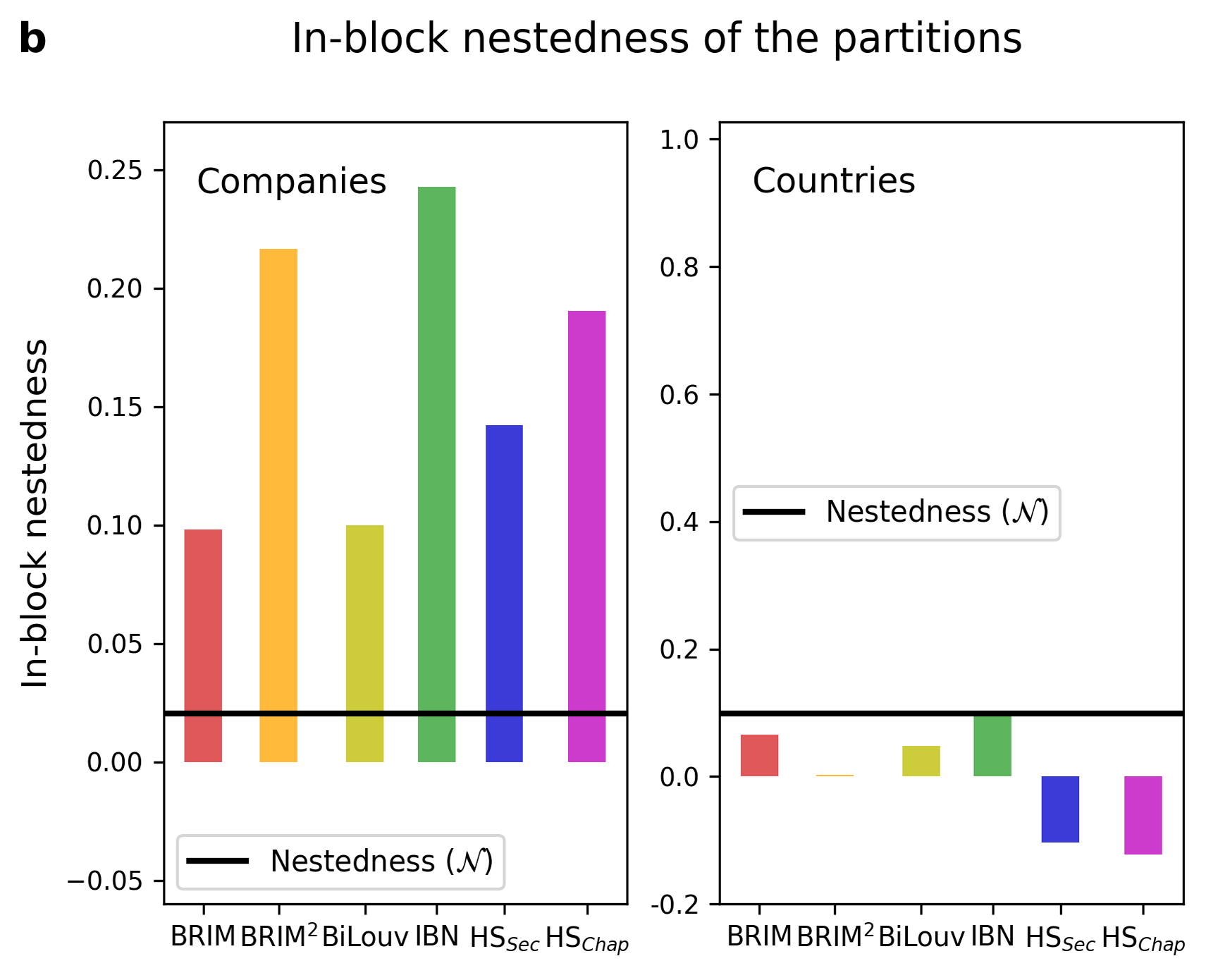}%
\caption{\textbf{Evaluating the statistical significance and the robustness of the in-block nestedness of the countries' and firm's ecosystems.} (a) Empirical values of the optimal degree of in-block nestedness, $\mathcal{I}^*$, and the global nestedness, $\mathcal{N}$, for both the firm-product and the country-product network; $\overline{\mathcal{I}^*}_{CM}$ and $\overline{\mathcal{N}}_{CM}$ denote the average of the two functions over 10 realizations of the randomised networks generated according to the bipartite configuration model. Differently from the country-product network, the firm-product network exhibits $\mathcal{I}^*/\mathcal{N}\gg1$, proving quantitatively the in-block nestedness of this system.  (b) Robustness analysis by using different partitions obtained by maximising the modularity function (BRIM, BRIM$^2$ and BiLouvain), or by maximising the $\mathcal{I}$ function and through the sector information (HS System). The value of the in-block nestedness $\mathcal{I}$ is always higher (lower) than the nestedness $\mathcal{N}$ in the case of companies (countries). 
}
\label{2}
\end{figure*}

Our findings quantitatively confirm the qualitative representation of Figs.~\ref{1}. We find a large $\mathcal{I}^{*}/\mathcal{N}$ ratio for firms ($\mathcal{I}^{*}/\mathcal{N}\simeq 12.0$; see Fig.~\ref{2}a), where the in-block nestedness maximisation produces a partition with more than 80 blocks, but not for countries ($\mathcal{I}^{*}/\mathcal{N}\simeq 1.02$), where only 2 modules are detected, of which the largest one includes the vast majority of the network nodes (97.6\%) - see Supplementary Fig. S5 for a visual representation. To rule out the possibility that large $\mathcal{I}^*$ values arise through random fluctuations~\cite{guimera2004modularity}, we compare the observed values of $\mathcal{I}^{*}$ against those obtained in randomised bipartite networks that preserve on average the nodes' degree (see Methods). We find that the in-block nestedness $\mathcal{I}^{*}$ of the firm-product network is significantly larger than that of the corresponding randomised network, whereas the same does not hold for the country-product network (Fig.~\ref{2}a), where the level of in-block nestedness is entirely due to the degree of global nestedness. Note that by testing the significance of this result with the BICM model, we are performing a highly conservative statistical validation, which can notoriously rule out global nestedness in most empirical networks~\cite{payrato2019breaking,bruno2020ambiguity}.
Taken together, these results demonstrate that nestedness is a global property for countries, while it emerges locally for firms.

This conclusion is robust with respect to alternative partitions of the network.
Specifically, the empirical result that $\mathcal{I}\gg \mathcal{N}$ holds not only for the optimal in-block nested partition ($\mathcal{I}=\mathcal{I}^*$), but also for reasonable alternative partitions determined by modularity maximisation (via the BRIM, BRIM$^2$ and BiLouvain methods) or economic sectors (based on HS sections, HS$_{\text{Sec}}$, and chapters, HS$_{\text{Chap}}$) -- see Methods for a summary of these partitioning methods. Although the value of $\mathcal{I}$ for these partitions is smaller than $\mathcal{I}^{*}$, it remains considerably larger than $\mathcal{N}$ (see Fig.~\ref{2}b) -- the fingerprint of a network where nestedness is a local network property, and not a global one. Remarkably, in the country-product network, none of the sub-optimal partitions achieves a value of $\mathcal{I}$ comparable to $\mathcal{N}$: we observe $\mathcal{I}\ll \mathcal{N}$ for all parititions but the optimal one (for which $\mathcal{I}=\mathcal{I}^*\simeq \mathcal{N}$; see Fig.~\ref{2}b).
This further confirms that nestedness is a global property of the country-product network.

\section*{Discussion}

Despite recent advances in Economic Complexity, comparing the structure and dynamics of economic ecosystems at the country and company scales remained elusive, mostly due to the scarcity of datasets on firms' export activities and the lack of specific methodologies. Here, we overcame this limitation by analysing a unique dataset of Italian firms' exports and a worldwide dataset of the export flows between countries, and by comparing the observed structure of the firms' and countries' ecosystems via recently introduced approaches~\cite{sole2018revealing,palazzi2021ecological}. 

Our results reveal that, when looking at an economic ecosystem at different scales, stark structural differences emerge. While we observed a globally nested structure at the country level, we found an in-block nested structure at firm level. We showed that the observed structural differences have profound implications for economic complexity rankings: the Fitness-Complexity algorithm~\cite{tacchella2012new} neglects the block structure of interactions and, as a result, it correctly extracts the economic value of products and the competitiveness of economic agents in the country-product network, but not in the firm-product network. 

Nevertheless, developing Economic-Complexity ranking and recommendation algorithms tailored to firms would have profound implications for managerial and policy-making decisions, innovation strategies and investments.
To this end, our findings suggest that the first crucial step should be the identification of the local ecosystem of the firms of interest and its boundaries. The appropriate context is not the entire network (as for countries), but is provided by the company-product blocks where the firms operate. Interestingly, since these local ecosystems are internally nested, then applying locally the Fitness-Complexity algorithm may still be an effective strategy to rank companies and products within their ecosystem. This analysis will be the subject of future works.


\section*{\label{sec:Methods}Methods}

\subsection*{\textbf{Data and network construction}}

We analysed two datasets: (1) the country-level dataset obtained from the UN-Comtrade dataset (https://comtrade.un.org), which is the standard database used in the Economic Complexity framework and (2) the ISTAT dataset concerning the export of Italian companies. In both datasets the export flows are recorded, and products are classified according to a six digit code which, after a data cleaning procedure, was standardised to the Harmonized System 1992 categorisation. We then coarse-grained the obtained classification by considering only the first 4 digits, resulting in a set of about 1,200 products.

The firms' dataset spans from 1993 to 2017 and it includes 879,280 companies. From year to year the number of companies exporting at least one product varies between 150,000 and 200,000. The countries' dataset spans from 1996 to 2018, and it includes 169 countries in total.

To perform a coherent analysis for both firms and countries we summed up the export volumes for all the available years and only kept the firms (countries) that remained active (for which data is available) for the entire time interval considered. As a result of this filtering procedure, a total of 18,349 firms and 161 countries were left. From these filtered data, we constructed the country-product and the firm-product bipartite binary networks. 

The criterion adopted in order to decide whether a country (company) can be considered or not as a competitive exporter of a particular product is the so-called Revealed Comparative Advantage (RCA) \cite{balassa1965trade}. For a pair $(i,\alpha)$ composed of a potential exporter $i$ (country or company) and a product $\alpha$, the RCA is defined in terms of the ratio between the fraction of export of product $\alpha$ by country (company) $i$ and the overall export of $\alpha$. The obtained quantity is then divided by the ratio between the total export of $i$ and the overall export by all countries (companies). This is the most natural way to remove trivial dependencies from the sizes of the economic agents and sectors. In formulas:
\begin{equation}
    RCA_{i\alpha} = \frac{\frac{q_{i\alpha}}{\sum_{i'}q_{i'\alpha}}}{\frac{\sum_{\alpha'}q_{i\alpha'}}{\sum_{i'\alpha'}q_{i'\alpha'}}}.
\end{equation}
As in previous works~\cite{hidalgo2009building,tacchella2012new}, a threshold value $R^*=1$ is used. As a result, a binary country (company)-product matrix $\textbf{M}$ is built, whose generic element is:
\begin{equation}
    M_{i\alpha} = \begin{cases} 1 & \mbox{if } RCA_{i\alpha}\geq R^*=1 \\ 0 & \mbox{if } RCA_{i\alpha}<R^*=1 \end{cases},
\end{equation}
i.e., country (company) $i$ can be considered a competitive exporter of product $\alpha$ if and only if $M_{i\alpha}=1$. In the equivalent network representation, the node of the country (company) $i$ is linked to the node of the product $\alpha$ if and only if $M_{i\alpha}=1$.
For the characterisation of the basic properties of the two constructed networks, see Supplementary Table S1.

\subsection*{Network analysis methods}

\paragraph*{\textbf{Modularity.}} 
We search for a (sub)optimal modular partition of the nodes by applying a variant of the BRIM (Bipartite, Recursively Induced Modules) algorithm\footnote{\url{https://github.com/genisott/pycondor}} \cite{platig2016bipartite} to maximise Barber’s modularity \cite{barber2007modularity}, defined as:
\begin{equation}
        Q = \frac{1}{E}\sum_{i=1}^{N_R}\sum_{\alpha=1}^{N_C}(M_{i\alpha}-P_{i\alpha})\delta(a_i,a_{\alpha}),
\end{equation}
where $E$ is the number of interactions (links) in the network, $M_{i\alpha}$ is the biadjacency matrix which denotes the existence of a link between row nodes $i$ and column nodes $\alpha$, $P_{i\alpha}=k_ik_{\alpha}/E$ is the probability that a link between nodes $i$ and $\alpha$ exists by chance under a degree-preserving null model, $a_i$ is a membership variable that defines the block to whom the node $i$ belongs, and $\delta(a_i,a_{\alpha})$ is the Kronecker delta function, which takes the value $1$ if nodes $i$ and $\alpha$ are in the same community, and 0 otherwise.

Given the resolution limit that affects modularity optimisation~\cite{fortunato2007resolution}, we also considered an alternative method that applies the BRIM algorithm twice, by performing community detection within the blocks identified through the first application of the algorithm. We refer to this method as BRIM$^2$.
In order to verify the robustness of the results, all the analyses were replicated using the BiLouvain algorithm, which is the extension to bipartite networks of the popular Louvain algorithm introduced by Blondel et al. \cite{blondel2008fast}.

\paragraph*{\textbf{Global and In-block nestedness.}}
In-block nested structures are patterns of interactions characterised by compartments of nodes that internally exhibit a nested pattern of interactions. 
Using the formulation developed in \cite{sole2018revealing}, the degree of in-block nestedness $\mathcal{I}$ of a network can be quantified as:
\begin{equation}
\label{eq:I}
\begin{aligned}
    \mathcal{I}=\frac{2}{N_R+N_C}\bigg{\{}\sum_{i,j}\frac{O_{i,j}-\langle O_{i,j} \rangle}{k_j(C_i-1)}\Theta(k_i-k_j)\delta(a_i,a_j) + \\\sum_{\alpha, \beta}\frac{O_{\alpha,\beta}-\langle O_{\alpha, \beta}\rangle}{k_{\beta}(C_{\beta}-1)}\Theta(k_{\alpha}-k_{\beta})\delta(a_{\alpha},a_{\beta})\bigg{\}},
\end{aligned}
\end{equation}
where $C_i$ is the number of nodes that belong to the block to whom the node $i$ belongs, $O_{i,j}$  measures the degree of links overlap between rows node pairs, $\langle O_{ij} \rangle$ represents the expected number of links between row nodes $i$ and $j$ in the null model and is equal to $\langle O_{ij} \rangle = k_ik_j/N_R$, and $\Theta(\cdot)$ is the Heaviside step function that guarantees that the overlap is computed only between pair of nodes such that $k_i > k_j$. The function $\mathcal{I}$, called in-block nestedness fitness, can be interpreted as a generalisation of the \textit{global nestedness} function:
\begin{equation}
\label{N}
\begin{aligned}
        \mathcal{N}=\frac{2}{N_R+N_C}\bigg{\{}\sum_{i,j}\frac{O_{ij}-\langle O_{i,j} \rangle}{k_j(N_R-1)}\Theta(k_i-k_j)+\\ \sum_{\alpha,\beta}\frac{O_{\alpha\beta}-\langle O_{\alpha, \beta}\rangle}{k_{\beta}(N_C-1)}\Theta(k_{\alpha}-k_{\beta})\bigg{\}},
\end{aligned}
\end{equation}
introduced in \cite{sole2018revealing} as an overlap-based metrics, inspired by the NODF (Nestedness metric based on Overlap and Decreasing Fill) \cite{almeida2008consistent}, which compares the observed level of nestedness with the expected value under a suitable null model. Noteworthy, the objective function $\mathcal{I}$ reduces to $\mathcal{N}$ if one considers a single block ($a_{i}=a_{\alpha}=a, \hspace{0.1in} \forall i, \alpha$).
Here we search for a (sub)optimal in-block nested partition of the nodes by applying a variant of the extremal optimisation algorithm \cite{duch2005community}, adapted to maximise the in-block nestedness function\footnote{\url{https://github.com/COSIN3-UOC/nestedness_modularity_in-block_nestedness_analysis}}. 

\paragraph*{\textbf{Null models and statistical tests.}} 
To statistically validate the degree of modularity and in-block nestedness, we have used the Bipartite Configuration Model (BiCM) \cite{squartini2011analytical, saracco2015randomizing} paired with the $p$-value.

The BiCM is an entropy-based and unbiased null model which preserves, on average, the degree of both rows and columns\footnote{\url{https://github.com/mat701/BiCM}}. 

The $p$-value is computed by measuring the frequency of matrices in the null ensemble that are more modular/in-block nested than the input matrix and a threshold value $\lambda=0.05$ is used to denote a statistically significant level ($p < \lambda$). For matrices where no randomised networks satisfy this condition, we conservatively assigned $p < 1/R$, where $R$ is the number of independently generated random matrices.

\paragraph*{\textbf{Sectoral partitions.}}
In addition to community detection methods based on maximising modularity and in-block nestedness, we also constructed partitions following the HS classification for products (Supplementary Sect. V). In particular, in one case (referred to as HS$_{Sec}$) we partitioned the products according to the 21 HS sections and then we assigned countries (companies) to the block corresponding to their highest export volume. The second method (referred to as HS$_{Chap}$) follows the same strategy, except that the product communities do not correspond to the 21 HS sections but to the 99 HS chapters.

\paragraph*{\textbf{Partition similarity measures.}}
To evaluate and compare the performances of the clustering algorithms, here we make use of similarity measures based on information theory, which are built on the idea that if two partitions are similar, one needs very little information to infer one partition given the other, and thus this extra information can be used as a measure of dissimilarity. In particular, we employ the so-called Adjusted Mutual Information (AMI) \cite{vinh2010information}, defined as:
\begin{equation}
    AMI = \frac{I(X, Y) - E\{I(X, Y)\}}{\frac{1}{2}[H(X)+H(Y)]-E\{I(X, Y)\}},
\end{equation}
where $X$ and $Y$ are two clusterings, $I(X, Y)$ is their mutual information and $E\{I(X, Y)\}$ is a correction for randomness using the permutation model \cite{lancaster}, in which clusterings are generated randomly subject to having a fixed number of clusters and points in each clusters.
Specifically, the AMI equals 1 when the two clusterings are identical, and 0 when the mutual information between the two clusterings equals its expected value.

\subsection*{Economic Complexity methods}

\paragraph*{\textbf{The Fitness-Complexity method.}}
The Fitness-Complexity method (FC) is a non-linear, iterative approach for Economic Complexity evaluation 
\cite{tacchella2012new}. 
Grounded on the nested network structure of the country-product network, the Fitness of a country $F_i$ is measured by the sum of its exported products, weighted by their Complexity $Q_{\alpha}$, while the Complexity of a product is measured in a nonlinear way. The underlying intuition is that the information that a product is made in some scarcely competitive countries is sufficient to conclude that the Complexity of such product is low. In formulas\footnote{\url{https://github.com/ganileni/ectools}} \cite{tacchella2012new}:
\begin{equation}
\begin{cases} \widetilde{F}_i^{(n)}=\sum_{\alpha} M_{i\alpha}Q_{\alpha}^{(n-1)} \\ \widetilde{Q}_{\alpha}^{(n)}= \frac{1}{\sum_i M_{i\alpha}\frac{1}{F_i^{(n-1)}}} \end{cases}
\longrightarrow 
\begin{cases} F_i^{(n)}=\frac{\widetilde{F}_i^{(n)}}{\langle \widetilde{F}_i^{(n)} \rangle_i} \\ Q_{\alpha}^{(n)}=\frac{\widetilde{Q}_{\alpha}^{(n)}}{\langle \widetilde{Q}_{\alpha}^{(n)} \rangle_{\alpha}} \end{cases}.
\label{FitCom}
\end{equation}
The initial conditions are $\widetilde{Q}_{\alpha}^{(0)}=1 \hspace{0.1in} \forall \alpha$ and $\widetilde{F}_i^{(0)}=1 \hspace{0.1in} \forall i$. The vector of country and product scores is the stationary point of these iterative equations. 
Noteworthy, this algorithm produces highly-nested biadjacency matrices \cite{pugliese2016convergence}. 

\paragraph*{\textbf{PRODY.}}The PRODY index \cite{hausmann2007you} is the weighted average of per capita GDPs $Y$, where the weights represent the revealed comparative advantage\footnote{The rational for using revealed comparative advantages as weights is to control for country size when ranking the products.} $R_{i\alpha}$ in product $\alpha$ for country $i$:
\begin{equation}
    PRODY_{\alpha} = \sum_{i}\frac{R_{i\alpha}Y_i}{\sum_{i}R_{i\alpha}}.
\end{equation}
A slight modification, known as $\log$PRODY and introduced in \cite{angelini2017complex}, consists in replacing GDPpc with its logarithm.
The reasoning behind this choice is that, since GDPpc’s of countries span about four orders of magnitude, the geometric mean is better suited to represent such a numeric distribution of values.

By construction, sectors with high values of ($\log$)PRODY are those where high income countries play a major role in world exports. Then, under the reasonable assumption that high income countries display a strong presence where comparative advantages are determined by factors such as know-how, technological skills and so on, sectors characterised by a high ($\log$)PRODY index are more sophisticated than sectors with a low value of the index.

\begin{acknowledgments}
M.S.M acknowledges financial support from the URPP Social Networks at the University of Zurich. All the authors acknowledge ISTAT for providing the firm-level data, and in particular Dr. Stefano Menghinello and Dr. Cristina Lanzi.
\end{acknowledgments}


\section*{COMPETING INTERESTS}
The authors declare no competing interests.


\bibliographystyle{apsrev4-2} 
\bibliography{apssamp}

\end{document}


\preprint{AIP/123-QED}

\title{Supplementary Information: The different structure of economic ecosystems at the scales of companies and countries}

\author{Dario Laudati}
\affiliation{\mbox{Dipartimento di Fisica, Sapienza Università di Roma, 00185, Rome, Italy}}
 \author{Manuel S. Mariani}
 \affiliation{Institute of Fundamental and Frontier Sciences, University of Electronic Science and Technology of China, Chengdu 610054, PR China}
\affiliation{URPP Social Networks, Universität Zürich, 8050 Zürich, Switzerland}
\author{Luciano Pietronero}%
\affiliation{%
 Centro Ricerche Enrico Fermi, 00184, Rome, Italy
}%
\author{Andrea Zaccaria}%
 \email{andrea.zaccaria@cnr.it}
\affiliation{%
\mbox{Istituto dei Sistemi Complessi, UOS Sapienza, CNR, 00185, Rome, Italy}
}%

\date{\today}

\keywords{Economic Complexity, network, ecosystems, nestedness, modularity, in-block nestedness}
\maketitle

\section{\label{sec:Charac}Network characterisation}
In Table \ref{tab:Properties} the fundamental quantities that characterise the company-product and the country-product networks are reported.
\begin{table}[!h]
\centering
\caption{General properties of the company-product and country-product networks.}
    \begin{tabular}{ccccccccc}
              General properties & Company-product & Country-product \\
    \hline
    \% of links validated by RCA & 32.4 & 15.6 \\
    Number of rows & 18,349 & 161 \\
    Number of columns & 1,233 & 1,242 \\
    Number of links & 288,586 & 29,432 \\
    Density & 0.0128 &  0.147 \\
    \hline
    Mean row degree & 15 & 182 \\
    Maximum row degree & 555 & 557 \\
    Minimum row degree & 1 & 4 \\
    \hline
    Mean column degree & 243 & 23 \\
    Maximum column degree & 2,729 & 78 \\
    Minimum column degree & 1 & 4 \\
    \hline
    \end{tabular}
\label{tab:Properties}
\end{table}\\
\newpage
\section{\label{sec:EFC}Fitness-Complexity algorithm}
\begin{figure*}[h!]
  \includegraphics[width=.49\linewidth]{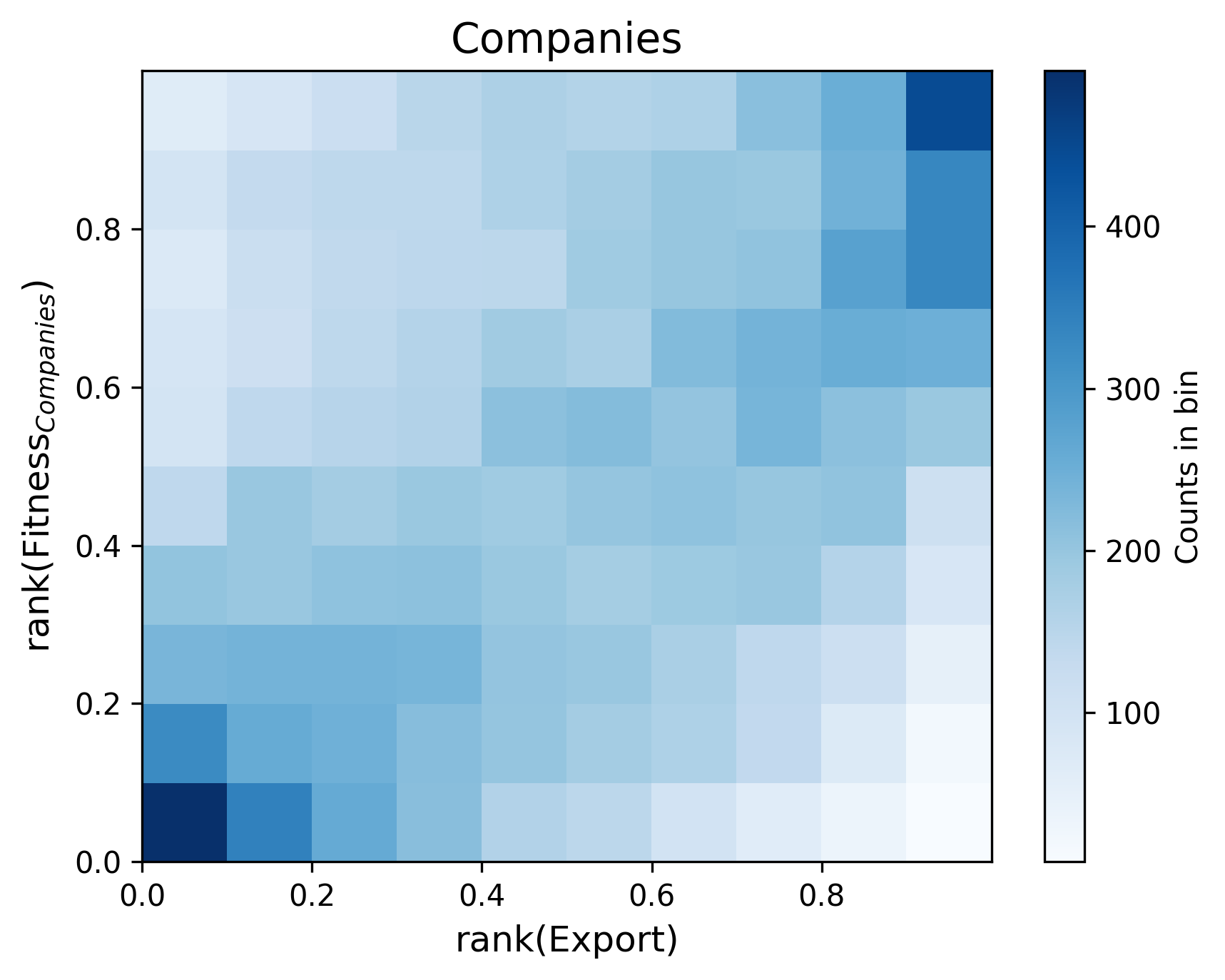}%
  \includegraphics[width=.49\linewidth]{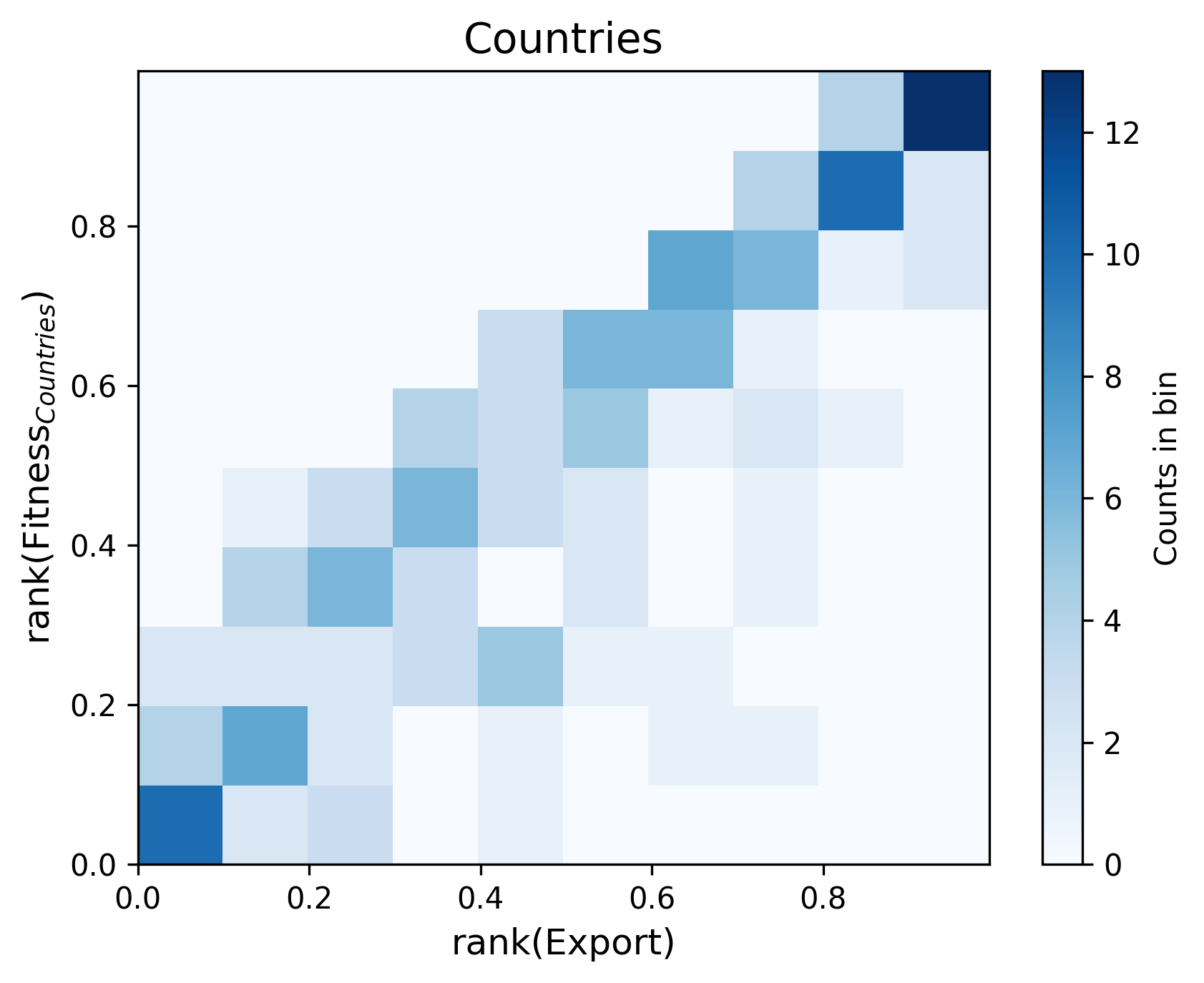}%
\caption{Comparison of the company and country rankings obtained with the Fitness-Complexity algorithms with those computed using the export volumes. Points are grouped in bins of size 0.1, with rankings
being normalised between 0 and 1. In the case of similar classifications, an accumulation of points around the secondary diagonal should be observed, which happens only at country level.}
\label{fig:hist_exp}
\end{figure*}

In the main text we have shown that, due to the marked modular structure that characterises the company-product network, the Fitness-Complexity algorithm [\onlinecite{tacchella2012new, cristelli2013measuring}] is able to capture the intrinsic sophistication of products only in the country-product network, but not in the company-product network. The fact that the predictions made on the network of companies are much weaker than those made on the network of countries is also evidenced by the correlation between Fitness and diversification, which is the basis of the Fitness-Complexity algorithm. This correlation, measured using the Spearman's correlation coefficient, is almost equal to 1 in the case of countries ($\rho=0.969$), while it is significantly smaller in the case of companies ($\rho=0.565$). \\
In order to confirm these considerations, we compare the Fitness, which is supposed to measure competitiveness, with the export volume, considered as a proxy for the performance of companies and countries (the higher the export volume, the higher the level of competitiveness). Given that export volume is an extensive index and there is no way to make it intensive (no other information about the companies is available), the intensive Fitness defined in Equation 7 in the main text is not suitable for this comparison. The corresponding extensive metrics is obtained by replacing the binary matrix $\textbf{M}$ with the weighted matrix $\textbf{W}$, whose elements $W_{cp}$ range from 0 to 1 and are defined as:
\begin{equation}
    W_{cp}=\frac{q_{cp}}{\sum_{c}q_{cp}},
\end{equation}
which means that the weight $W_{cp}$ is the fraction of export of product $p$ held by country (company) $c$.\\
Figure \ref{fig:hist_exp} shows the 2d histograms comparing the rankings obtained with the Fitness-Complexity algorithms with those computed using the export volume. Although there may not be a complete correspondence, since one quantifies the current power while the other also the ``potential” (capabilities), these two quantities are still expected to be significantly correlated.
\\
As a matter of fact, a significant correlation is present only in the case of countries ($\rho=0.887$), while for companies it seems that Fitness is not able to quantify the level of competitiveness ($\rho=0.378$).\\
\newpage

\section{\label{sec:Mod}Modularity}
\begin{figure*}[h!]
  \includegraphics[width=.48\linewidth]{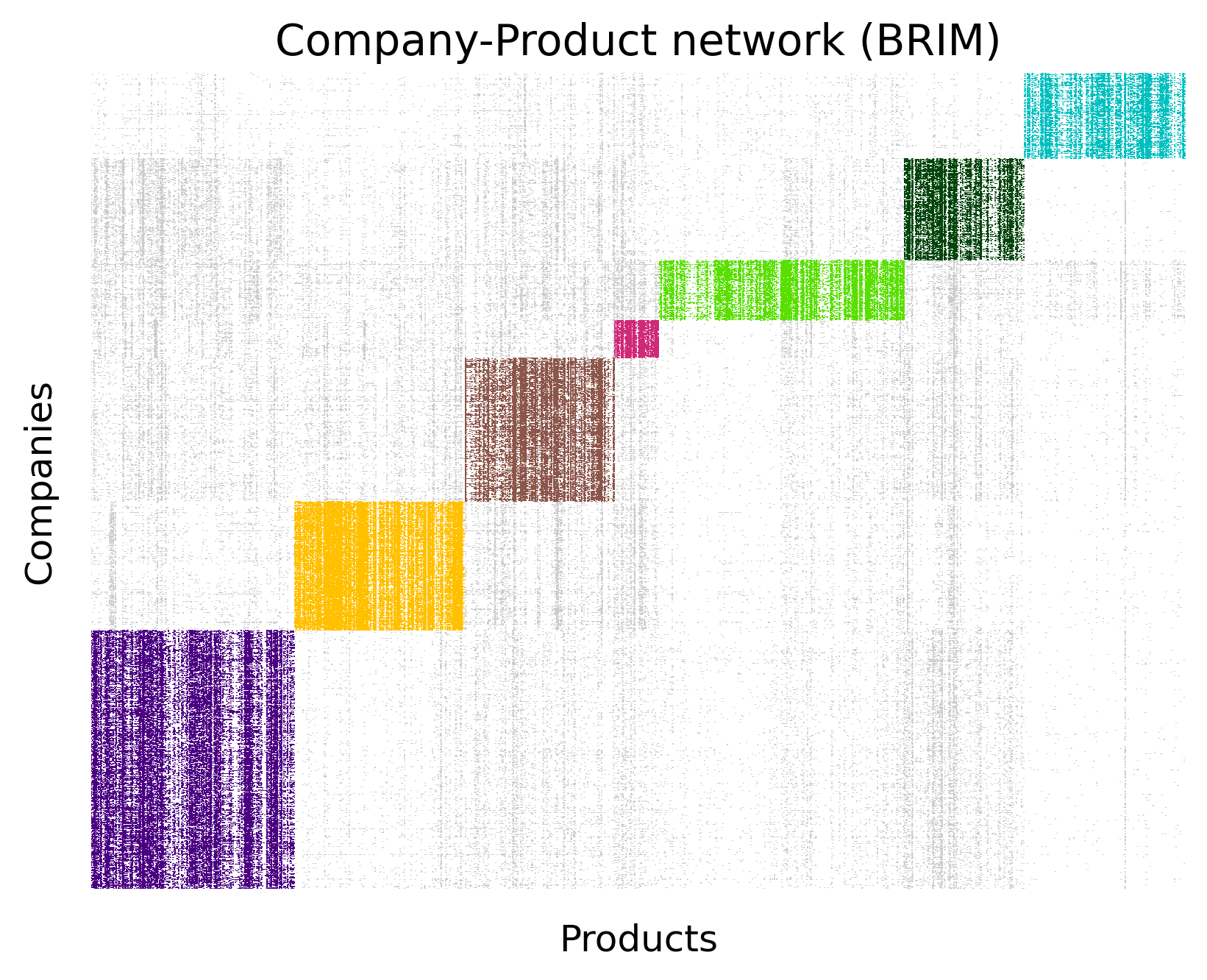}%
  \includegraphics[width=.48\linewidth]{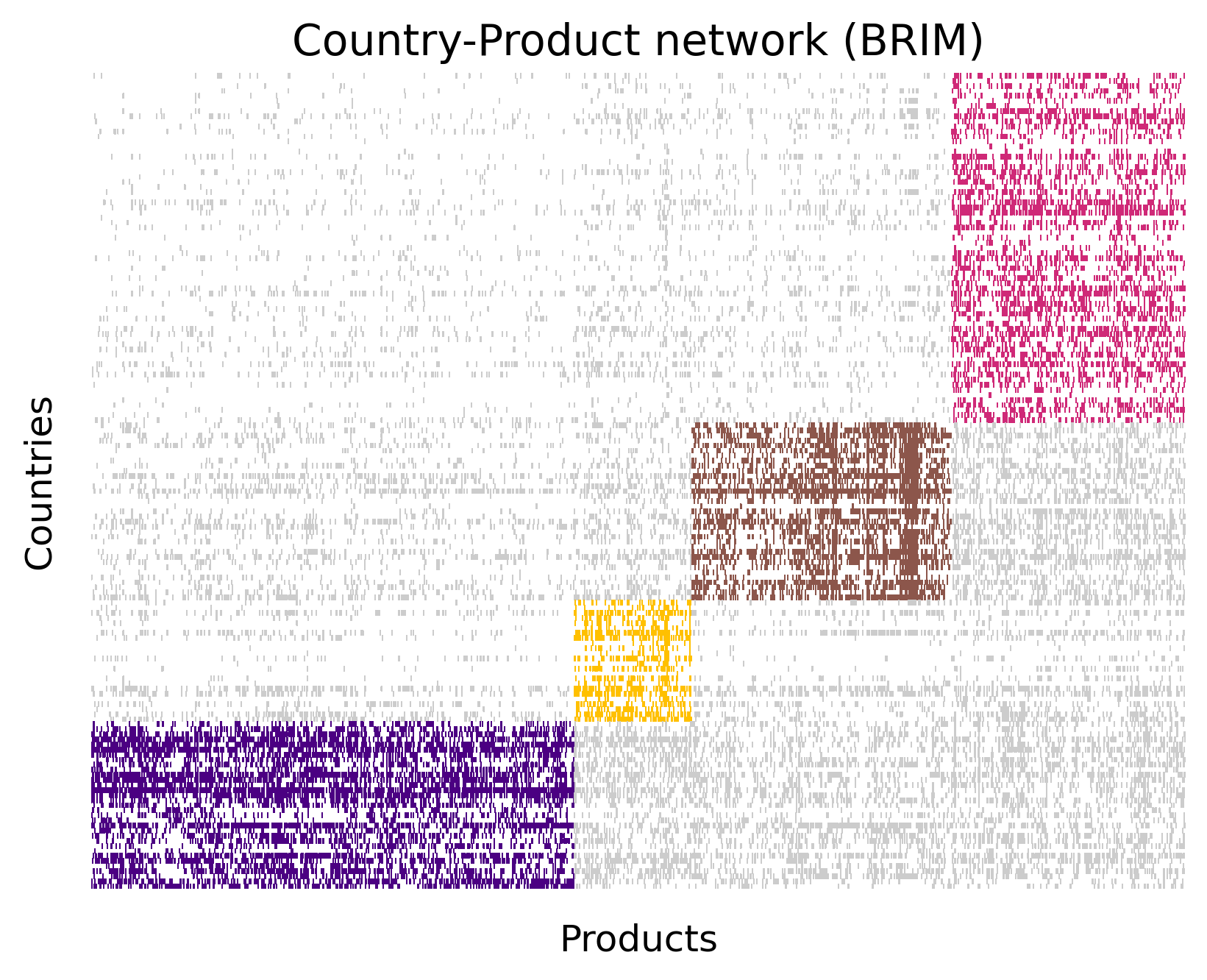}%
\caption{Company-product and country-product networks, where the communities detected by the BRIM algorithm [\onlinecite{platig2016bipartite}] are highlighted. Only for companies a clear division into modules emerges.}
\label{fig-Mod}
\end{figure*}

Figure \ref{fig-Mod} shows the company-product and the country-product networks, in which rows and columns are rearranged to highlight the modularity-based communities. While in the case of the companies a marked modular structure emerges (the modules contain more than 70\% of the links), for countries the partitions are very noisy and it is hard to distinguish clear blocks (as they include only 50\% of the total connections).

\subsection{Comparison with HS Sections}
\begin{figure*}[h!]
  \includegraphics[width=.48\linewidth]{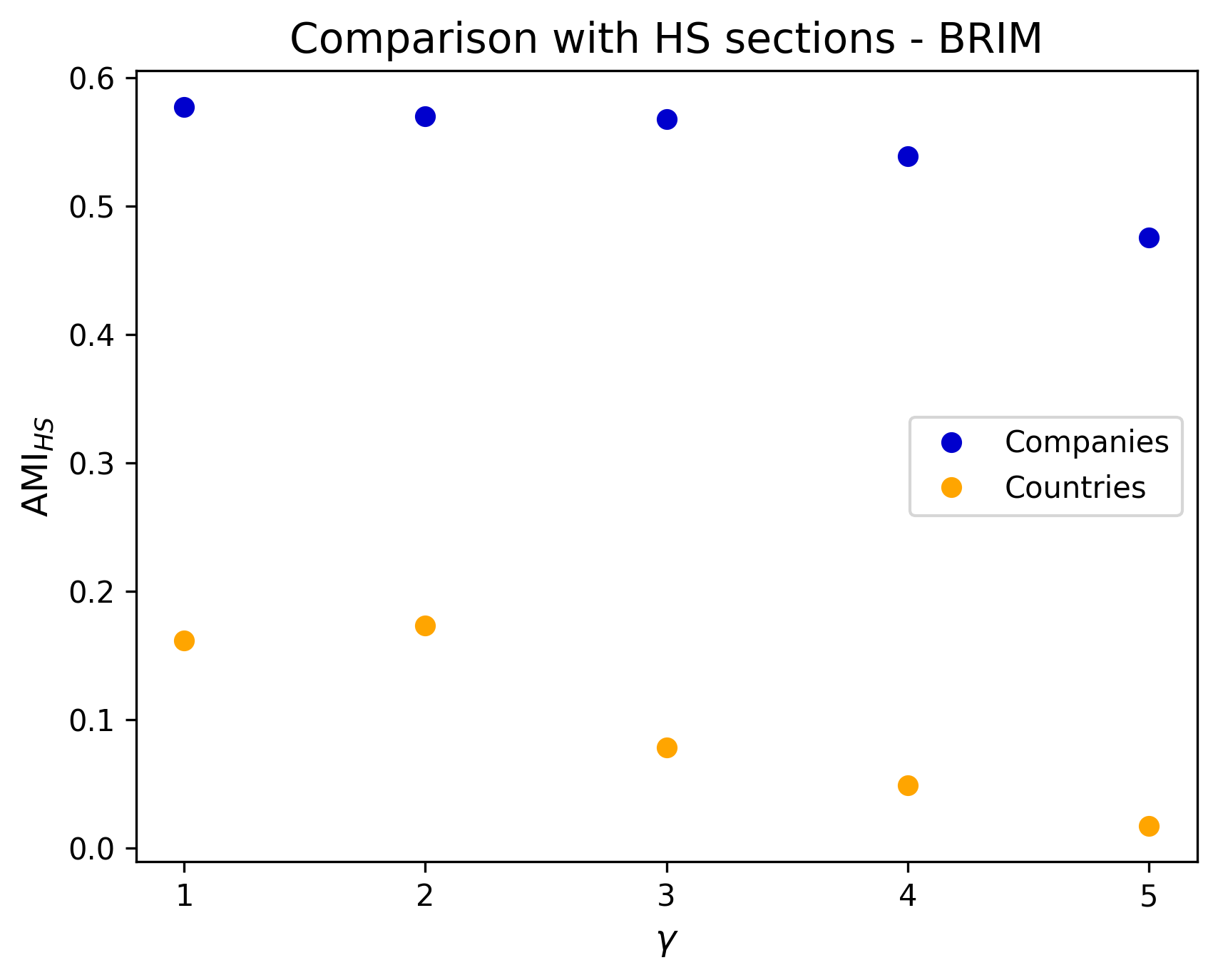}%
  \includegraphics[width=.48\linewidth]{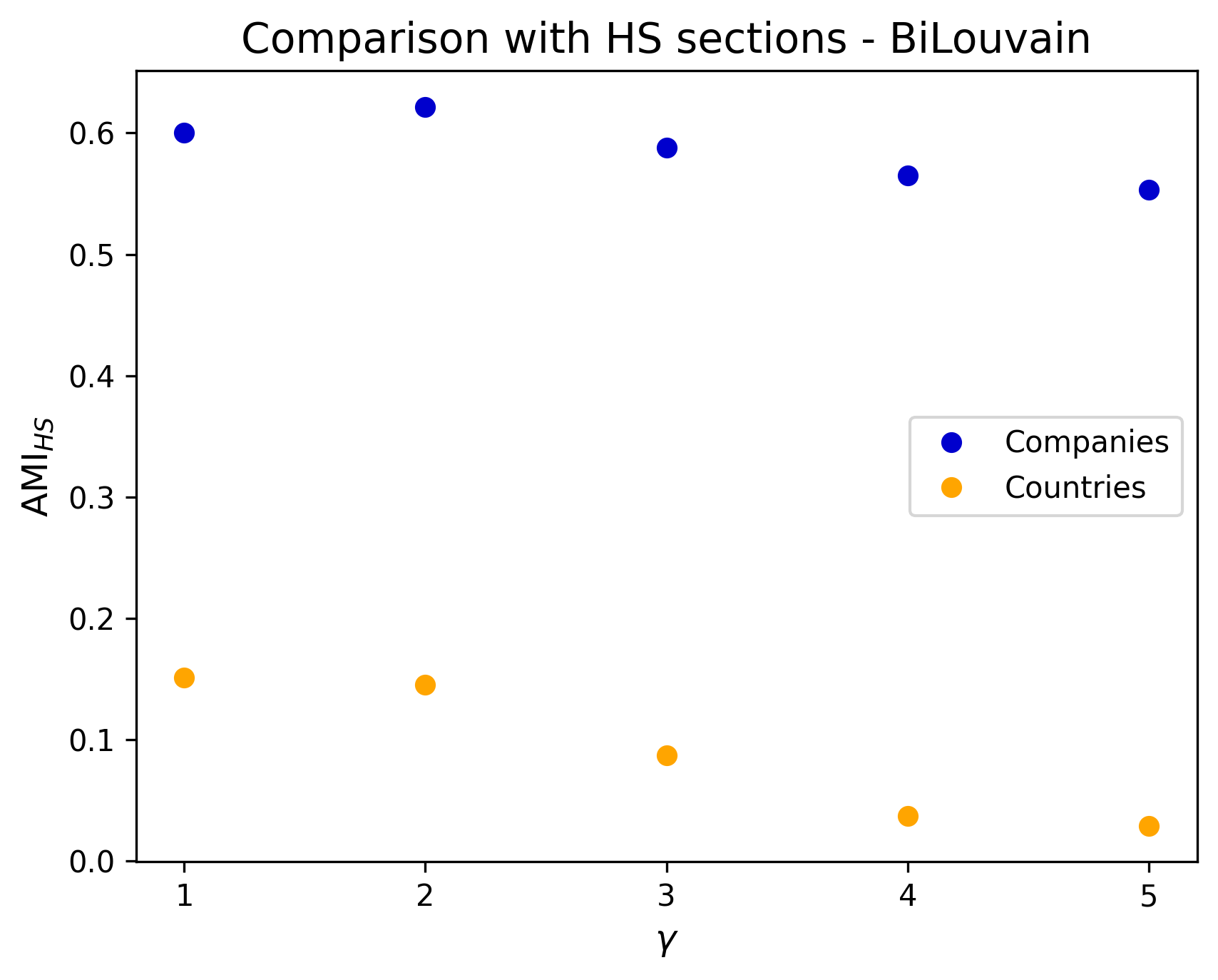}%
\caption{Comparison between HS sections and the partitions identified by two different modularity maximisation algorithms as the resolution parameter $\gamma$ varies. Only for firms the partitions are significantly correlated.}
\label{fig:ami_res}
\end{figure*}

In order to interpret the blocks detected by modularity optimisation, in the main text we have investigated their sector composition, through a comparison with the partitions corresponding to the 21 HS sections. Given that the specificity of the modules can affect the outcome of the comparison, here we show the results of several analysis (see Figure \ref{fig:ami_res}), carried out varying the resolution parameter $\gamma$, which allows to tune the characteristic size of the blocks and, therefore, to uncover modules at different scales [\onlinecite{fortunato2007resolution, reichardt2006statistical}].

As a matter of fact, we obtain a significant level of similarity only in the case of firms, while for countries there is no agreement with the HS classification for any value of the resolution parameter. Noteworthy, this result remains valid also if the comparison is carried out with the 99 HS chapters instead of the 21 sections.

\subsection{Revealing the mistery of modularity for countries}
\label{Boh}
\begin{figure*}[h!]
  \includegraphics[width=.48\linewidth]{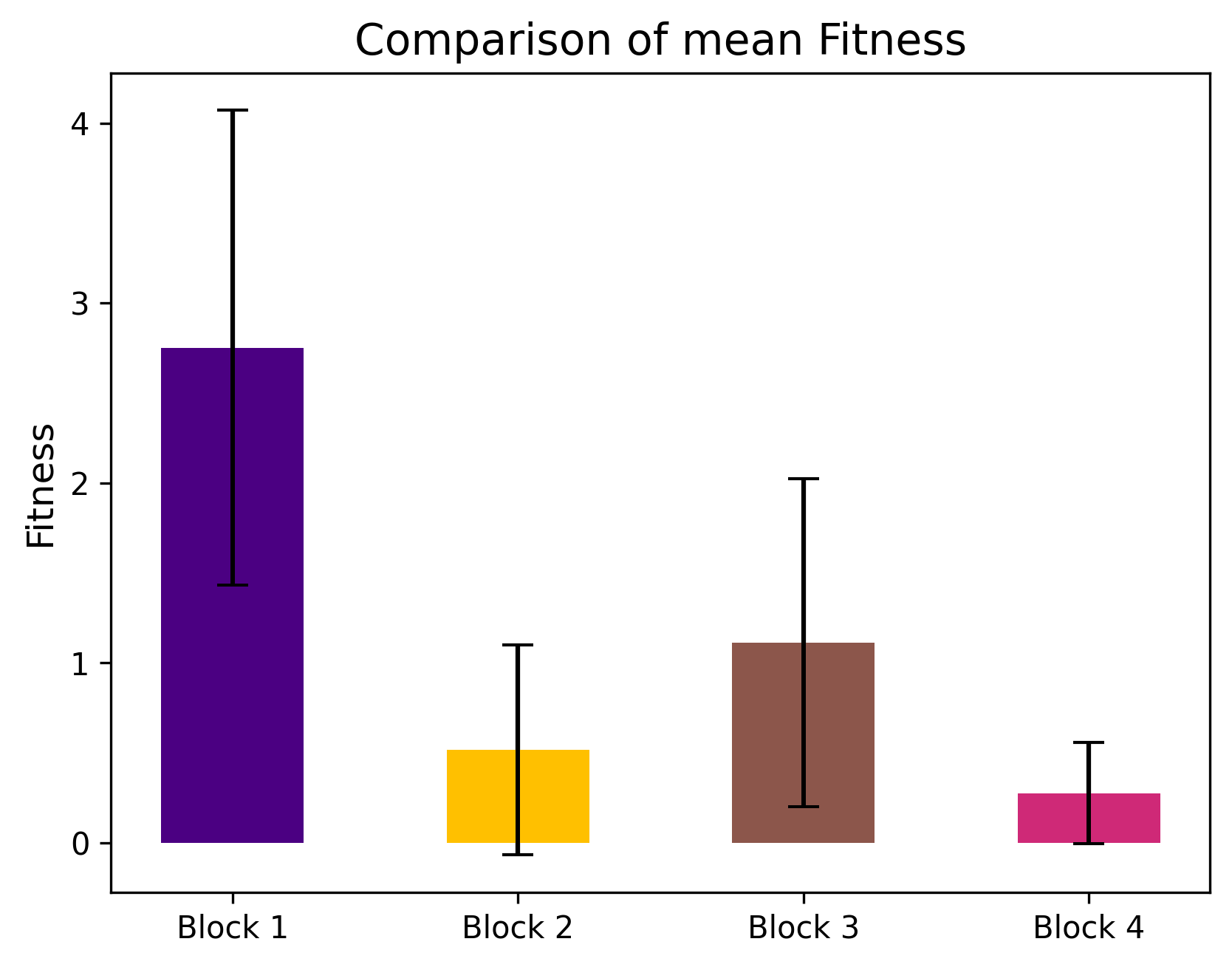}%
  \includegraphics[width=.48\linewidth]{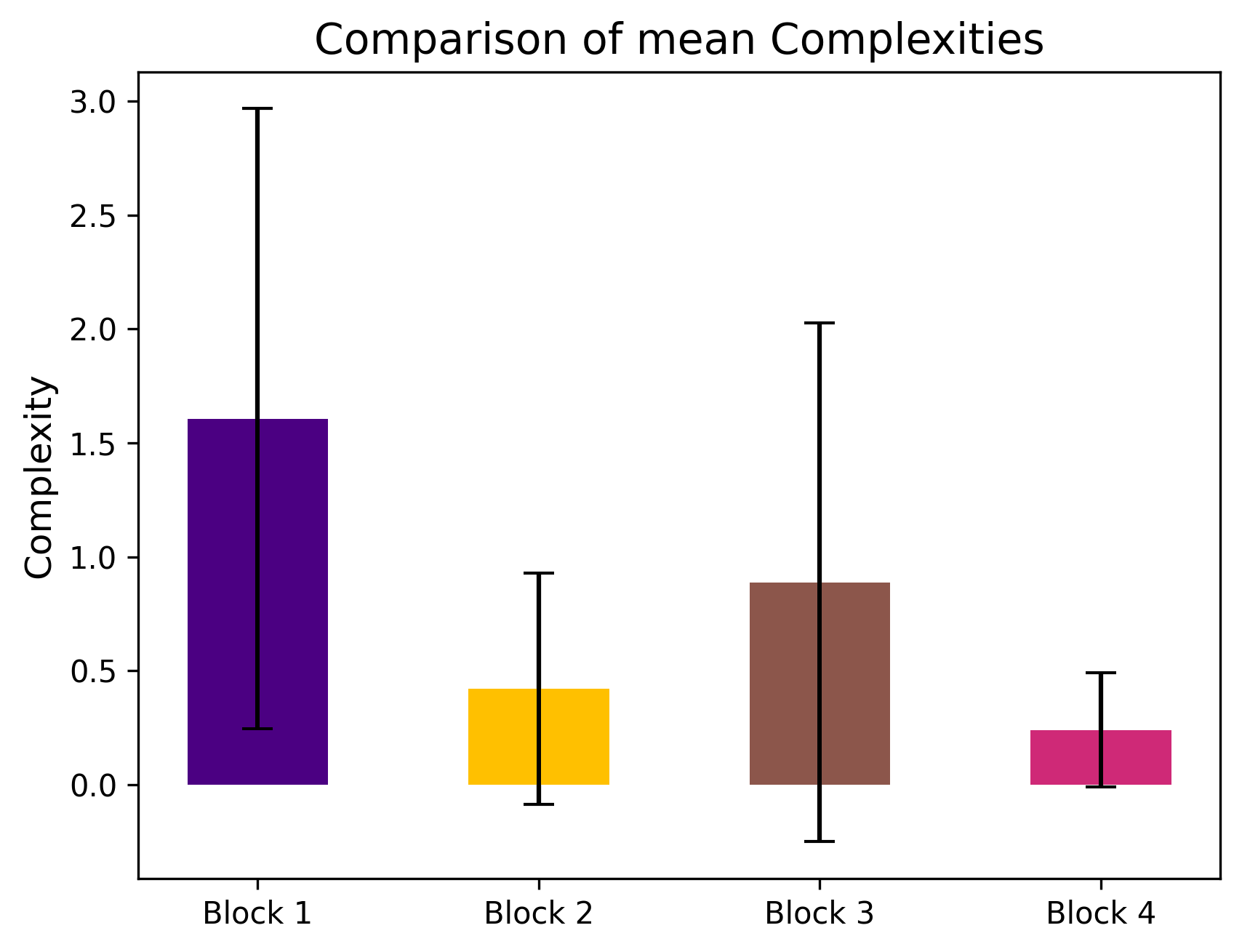}%
\caption{Comparison of the average Fitness and Complexity of the blocks identified by the BRIM algorithm in the country-product network.}
\label{fig:modules}
\end{figure*}

Figure \ref{fig:modules} shows the average Fitness and Complexity of each detected block of the country-product network, where the error bar corresponds to one standard deviation.

The four modules correspond, on average, to very different levels of Fitness and Complexity. Moreover, a significant one-to-one correspondence between the blocks of the two types clearly emerges: the module with higher Fitness is associated with the module with higher Complexity, and so on. Noteworthy, the standard deviation for each module assumes very high values, underlining a marked heterogeneity (diversification).\\
These results indicate that the modules that are identified in the country-product network are driven by the different degree of country diversification and product ubiquity rather than by an actual specialisation in production. In other words, for countries the communities of products are associated to the industrialisation level of the related exporters: the blocks are not made of homogeneous products but by those that can be efficiently exported by countries with strong industrial capabilities.

\subsection{Robustness check}
\begin{figure}[h!]
\centering
\includegraphics[width=4in]{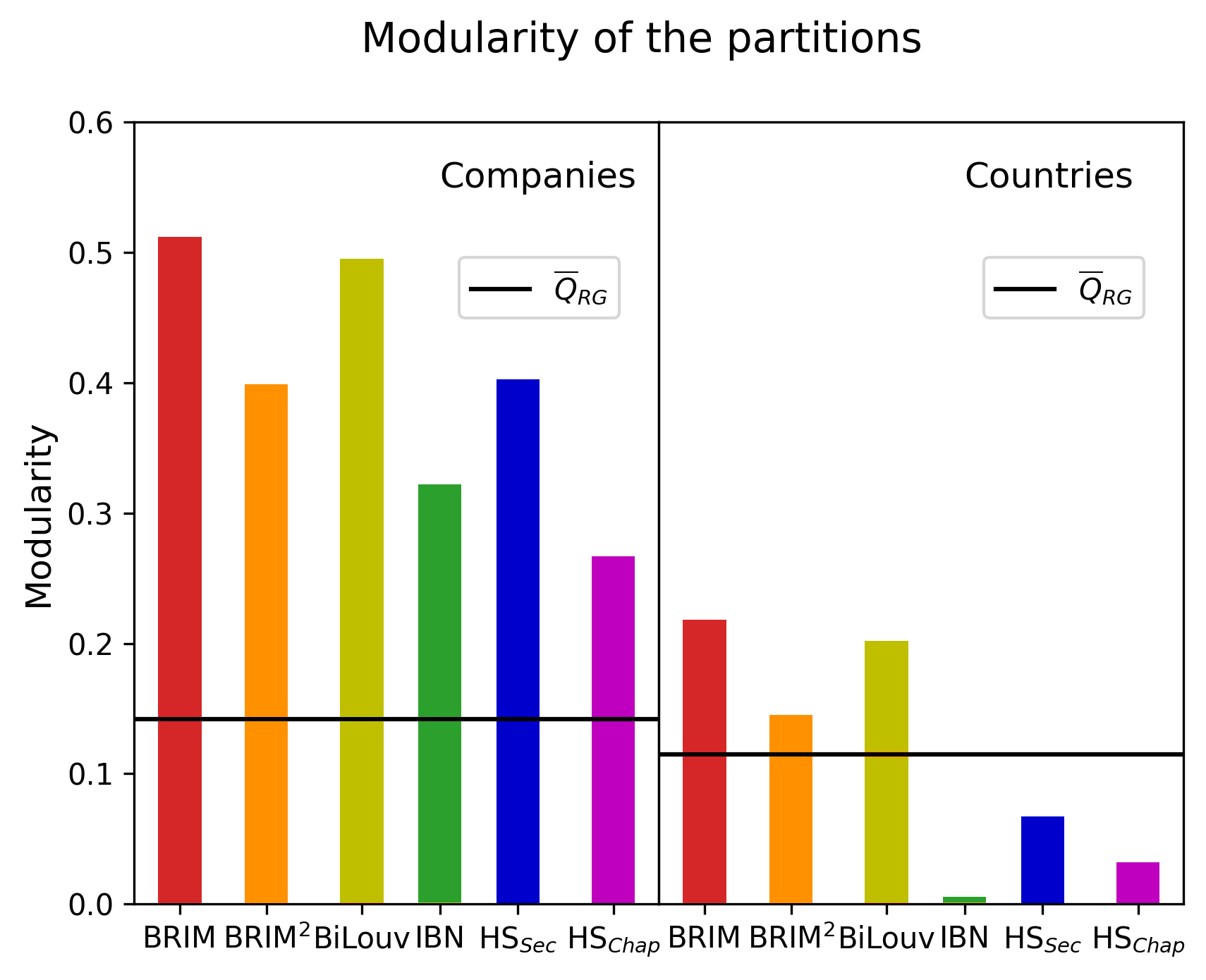}
\caption{Value of the modularity function for different partitions, obtained through modularity maximisation, by maximising the in-block nestedness and through the sector information (HS System). Only for companies modularity is persistent on all partitions.}
\label{fig-Q_Part}
\end{figure}

Having highlighted the structural differences between the country-product and the firm-product networks, we now check the robustness of the results. To this end, we use several community detection strategies, based on both modularity and in-block nestedness optimisation and on sectoral partitions (HS System) - see Methods in the main text.\\
We then measured and compared the modularity of each partition with the mean modularity over an ensemble of randomised networks [\onlinecite{straka2018ecology}], so as to assess the quality of each divisive strategy.\\
The obtained results are presented in Figure \ref{fig-Q_Part}.
\newpage
In the case of companies, we observe a very high level of modularity for all the partitions identified, while the same does not hold true for countries. In particular, the very low degree of modularity observed for the HS$_{Sec}$ and HS$_{Chap}$ methods indicates that it is not possible to associate a country with a single production sector, which again underlines that countries are not specialised entities.

\newpage
\section{In-block nestedness}
\begin{figure*}[h!]
  \includegraphics[width=.49\linewidth]{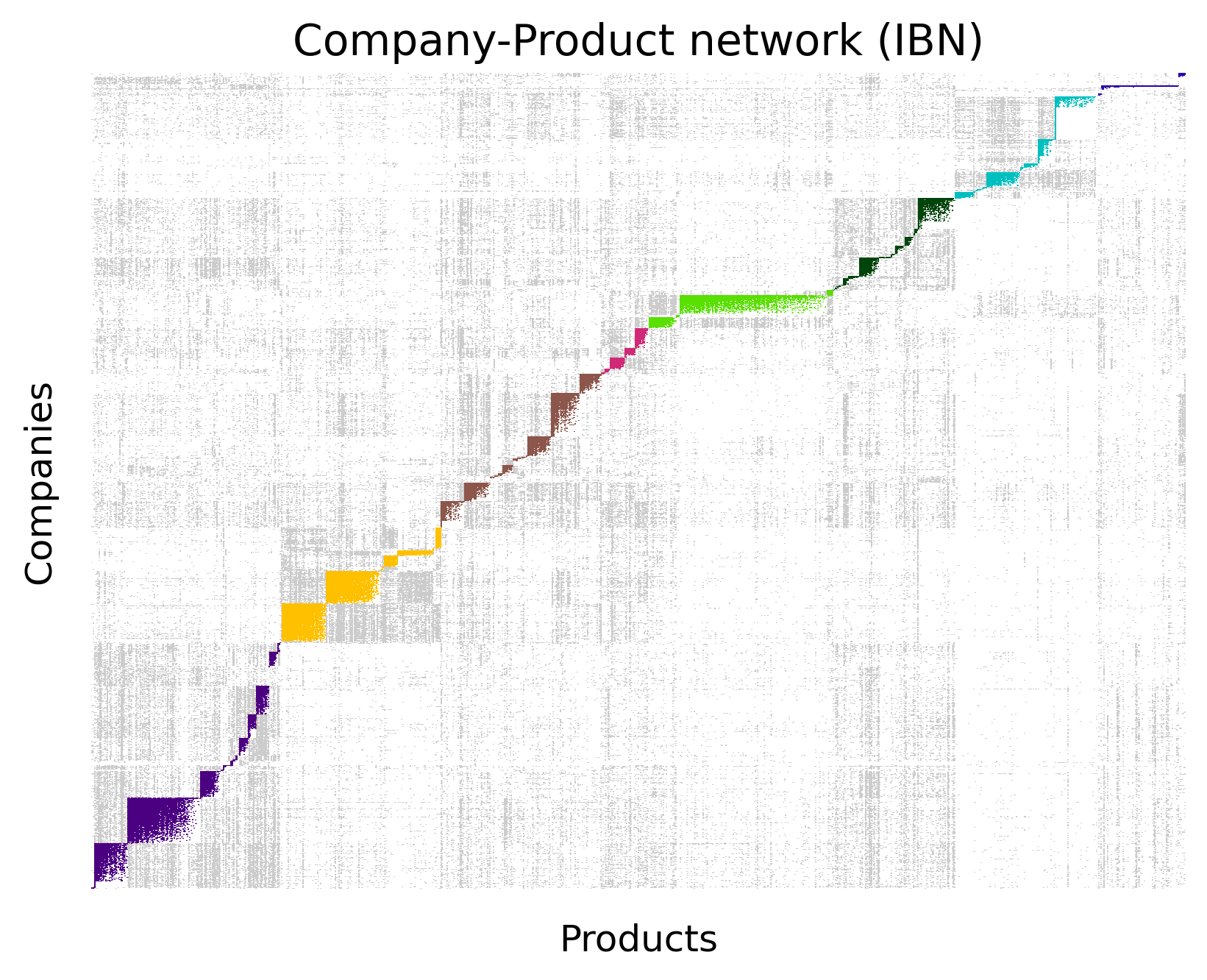}%
  \includegraphics[width=.49\linewidth]{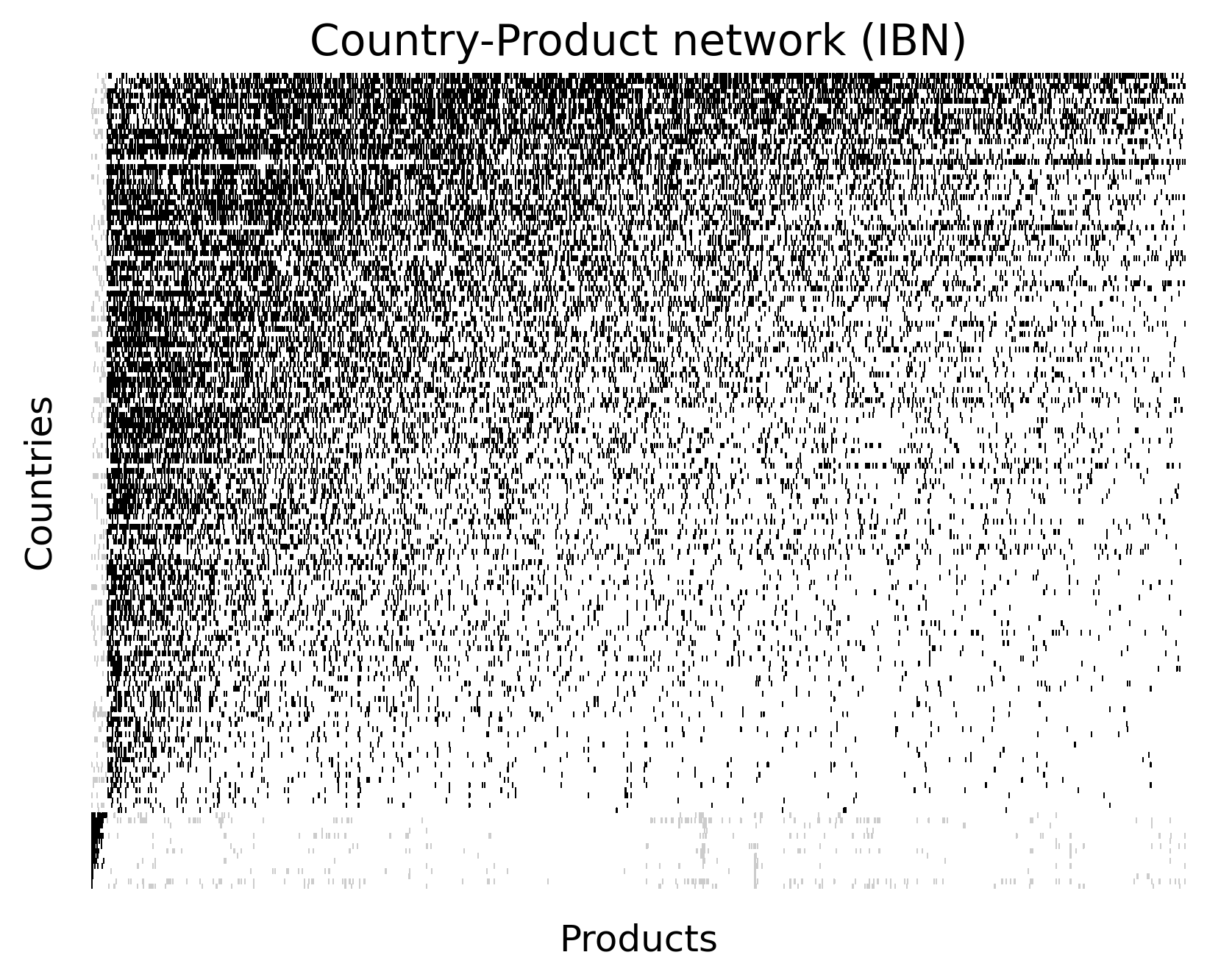}%
\caption{Company-product and country-product networks, where the communities detected through the optimisation of the in-block nestedness function are highlighted. While for countries nestedness emerges as a global property (right panel), for companies it arises locally (left panel), within subsets of similar firms producing the same type of products (in line with with exogenous classifications).}
\label{fig:ibn}
\end{figure*}

Figure \ref{fig:ibn} shows the company-product and country-product matrices, in which rows and columns are rearranged to highlight the communities obtained through the maximisation of the in-block nestedness fitness [\onlinecite{sole2018revealing, mariani2021absence}].

\begin{figure}[h!]
    \centering
    \includegraphics[width=0.49\linewidth]{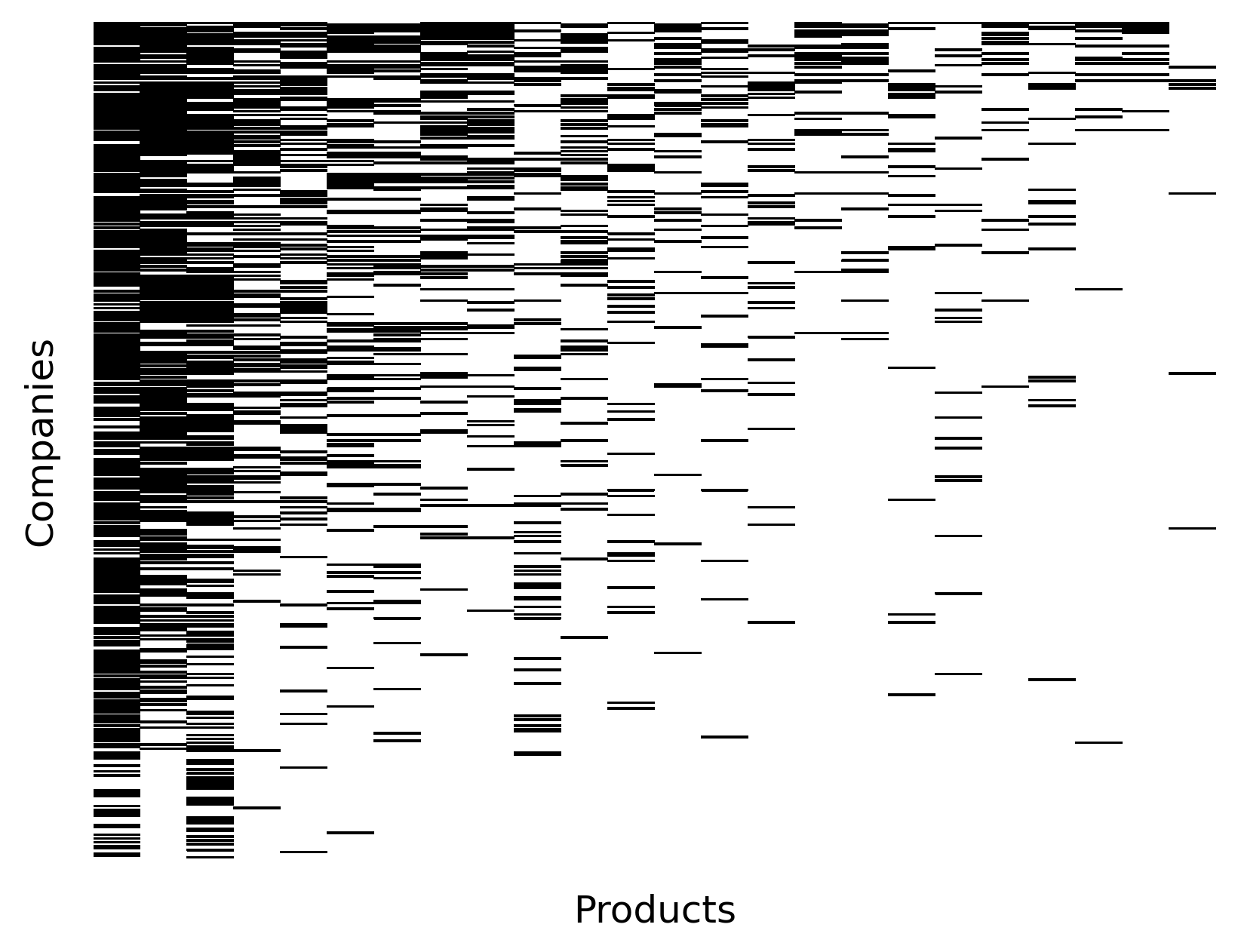}
    \includegraphics[width=0.49\linewidth]{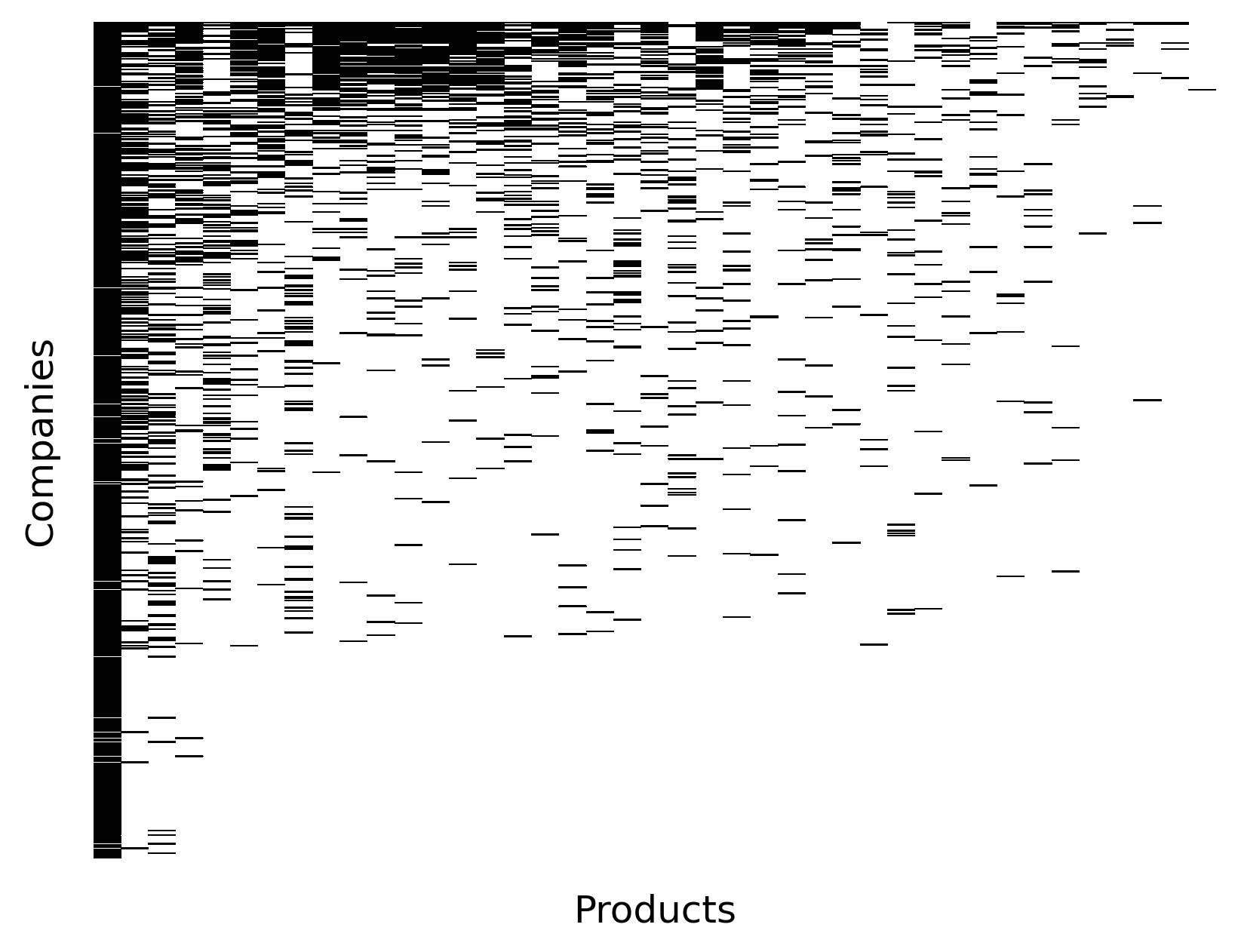}
    \includegraphics[width=0.49\linewidth]{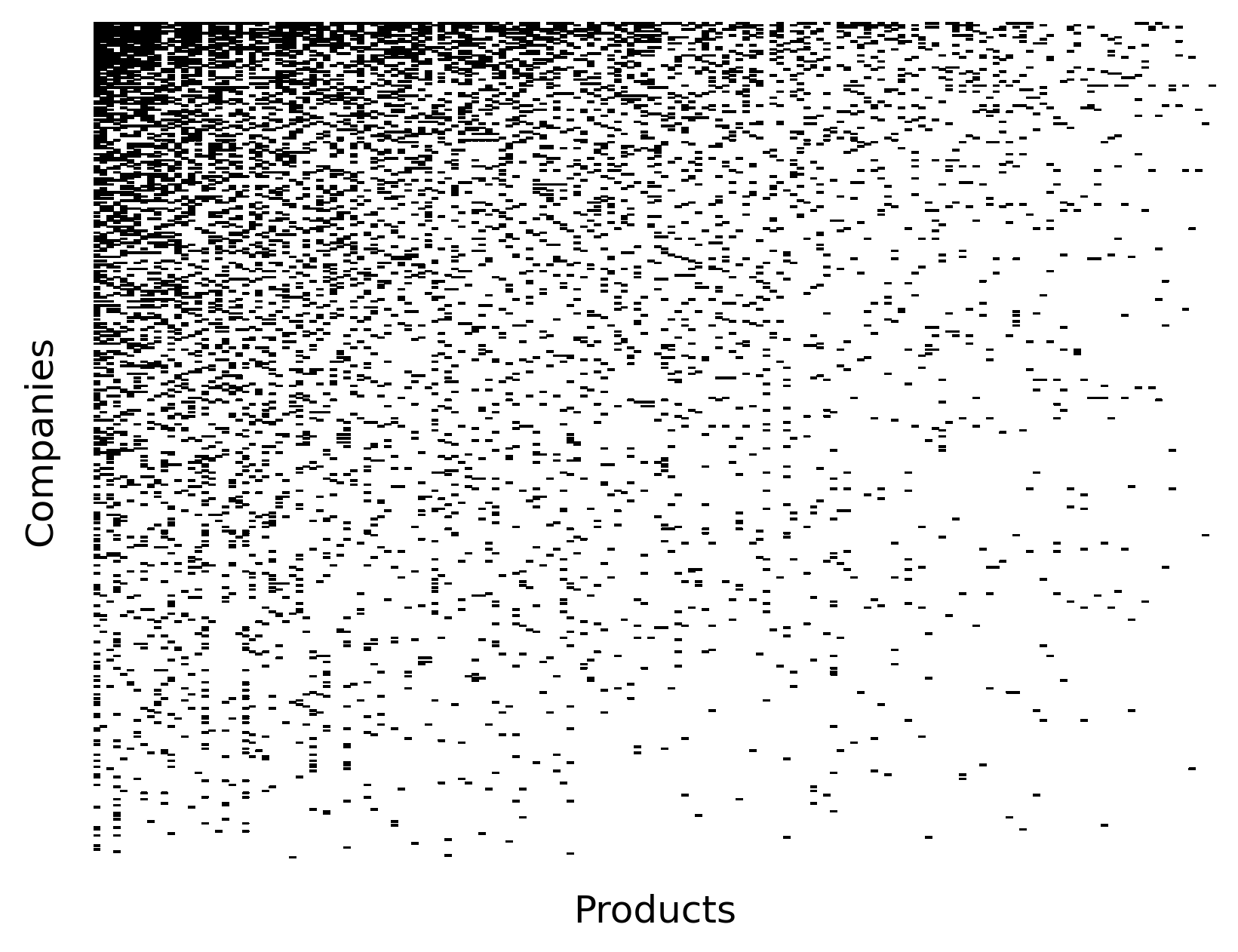}
    \includegraphics[width=0.49\linewidth]{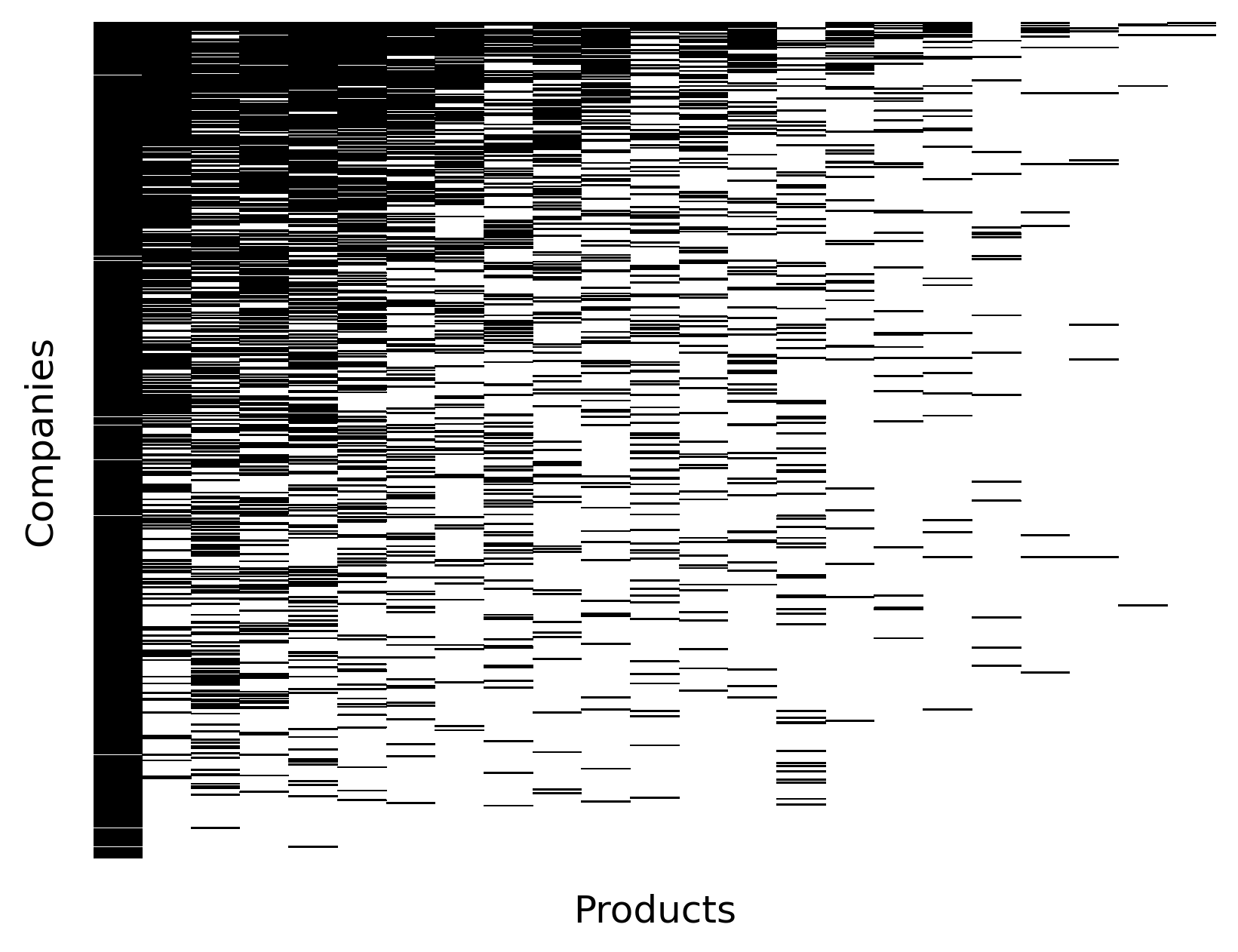}
    \caption{Zoom in on some of the modules identified by the in-block nestedness optimisation procedure in the company-product network. Top left panel: the products that constitute the module belong to the jewellery industry. Top right panel: the products belong to the iron and steel business. Bottom left panel: the products correspond to organic and inorganic chemicals. Bottom right panel: the products belong to the machinery sector.}
    \label{fig:Mod_IBN}
\end{figure}

As already pointed out in the main text, the in-block nestedness maximisation produces a partition with more than 80 blocks for firm, while only 2 modules are detected for countries, of which the largest one includes the vast majority\footnote{The smaller block is made up of a group of 15 countries, mainly belonging to the Middle East (among which Qatar, Kuwait, Oman and United Arab Emirates), and a set of products belonging to the area of mineral fuels and oil resources.} of the network nodes (97.6\%). As a result, it is safe to conclude that for countries the level of observed nestedness is a consequence of the global nested structure of the matrix, while for companies it is due to the emerge of many local nested structures.\\
Noteworthy, although the in-block nestedness and modularity functions are very different (one is based on the overlaps between nodes that belong to the same cluster while the other on link density), for companies their optimisation leads to partitions that are quite similar to each other (AMI$=0.496$). Interestingly, this correlation increases significantly when modularity is optimised iteratively (BRIM$^2$ method, AMI$=0.602$), suggesting that in-block nestedness optimization is also able to detect smaller blocks that are instead merged in modularity maximisation and are only revealed through an iterative operation. Moreover, the IBN partitions detected in the company-product network are also quite similar to sectoral partitions (HS sections, AMI$=0.484$), implying that the identified blocks are mostly composed by the same kind of products. Overall, it can be concluded that these firm-product modules consist of products that are very specific (more than 80 communities) but still similar to each other, in line with the divisions by sector. As a result of this specificity, the modules contain only 36\% of the total links (as opposed to 70\% in BRIM and 44\% in BRIM$^2$) and many “secondary blocks” are formed around the diagonal. It is important to note, however, that the optimisation of in-block nestedness does not aim to form “traditional” communities characterised by high internal density and few external links, but seeks communities to maximise nestedness.\\
For completeness, Figure \ref{fig:Mod_IBN} shows some of the largest blocks identified by in-block nestedness maximisation in the company-product network, where rows and columns are reordered according to the degree.\\
As expected, all the blocks exhibit a clear nested structure.
\newpage

\newpage

\section{\label{sec:HS}The Harmonized System}
The Harmonized System (HS) is an international nomenclature for the classification of products, which enables all physical goods moving across borders to be assigned to a class in a uniform manner all over the world.\\
Introduced in 1988, during the years it has undergone several changes, called revisions, which entered into force in 1996, 2002, 2007, 2012 and 2017.\\ 
It is used, as of December 2018, as the basis for Customs tariffs and for the compilation of international trade statistics, by more than 200 economies and Customs or Economic Unions (of which 157 are Contracting Parties to the HS Convention).\\
The HS comprises approximately 5,300 article/product descriptions that appear as headings and subheadings, arranged in 99 chapters, grouped in 21 sections.\\
\begin{table}[!h]
\centering
\caption{Classification of products in the 21 HS sections}
\resizebox{0.9\textwidth}{!}{%
    \begin{tabular}{ccc}
              & First two digits & HS Section \\
    \hline
     & 01-05 & Live animals; animal products \\
     & 06-14 & Vegetable products \\
     & 15 & Animal or vegetable fats and oil and their cleavage products \\
     & 16-24 & Prepared foodstuffs; beverages, spirits and vinegar; tobacco \\
     & 25-27 & Mineral products \\
     & 28-38 & Products of the chemical or allied industries \\
     & 39-40 & Plastics, rubber and articles thereof \\
     & 41-43 & Raw hides and skins, leather and furskins; travel goods \\
     & 44-46 & Wood, cork and articles thereof; manufactures of plaiting materials \\
     & 47-49 & Pulp of wood, paper and paperboard \\
     & 50-63 & Textile and textile articles \\
     & 64-67 & Footwear, headgear, umbrellas, walking-sticks and prepared feathers \\
     & 68-70 & Articles of stone, plaster and cement; ceramics; glass \\
     & 71 & Pearls, precious stones, imitation jewellery and coin \\
     & 72-83 & Base metals and articles of base metal \\
     & 84-85 & Machinery and mechanical appliances; electrical equipment \\
     & 86-89 & Vehicles, aircraft, vessels \\
     & 90-92 & Optical, measuring, precision and medical instruments \\
     & 93 & Arms and ammunition \\
     & 94-96 & Miscellaneous manufactured articles \\
     & 97 & Works of art, collectors' pieces and antiques \\
    \hline
    \end{tabular}%
     }
\label{tab:HS}
\end{table}\\
Each product is identified by a six digits code, which can be broken down into three parts:
\begin{itemize}
    \item the first two digits (HS-2) identify the chapter;
    \item the next two digits (HS-4) identify headings;
    \item the last two digits (HS-6) identify the sub-headings.
\end{itemize}
As the number of digits increases, the detail level of the product description increases. For instance, HS code 100630 belongs to Section II (Vegetable products) and consists of Chapter 10 (Cereals), Heading 06 (Rice), and Subheading 30 (semi-milled or wholly milled, whether or not polished or glazed).

\newpage
\nocite{*}
\bibliographystyle{apsrev4-2} 
\bibliography{Supplementary}